\documentclass{aa}  
\usepackage{graphicx}
\usepackage{natbib}
\usepackage{epsf}
\usepackage{psfig}
\usepackage{lscape}
\usepackage{txfonts}
\usepackage{balance}
\begin{document}
   \title{Deep multi-frequency rotation measure tomography of the galaxy cluster A2255\thanks{Data cubes are only available in electronic form at the CDS via anonymous ftp to cdsarc.u-strasbg.fr (130.79.128.5) or via http://cdsweb.u-strasbg.fr/cgi-bin/qcat?J/A+A/}}

%   \subtitle{I. Overviewing the $\kappa$-mechanism}

   \author{R.~F.~Pizzo\inst{1}, A.G. de Bruyn\inst{1,2}, G. Bernardi\inst{1,3}, and M.~A. Brentjens\inst{2}}
%  \author{}

     \institute{Kapteyn Institute, Postbus 800, 9700 AV Groningen, The
         Netherlands \\ \email{pizzo@astro.rug.nl} 
     \and 
         ASTRON, Postbus 2, 7990 AA Dwingeloo, The Netherlands 
         \and 
         Harvard-Smithsonian Center for Astrophysics, 60 Garden Street, Cambridge, MA 02138, USA \\ 
         \\}

%   \date{Received September 15, 1996; accepted March 16, 1997}

%\abstract{}{}{}{}{} 
% 5 {} token are mandatory
 
  \abstract{}{By studying the polarimetric properties of the radio
galaxies and the radio filaments belonging to the galaxy cluster \object{Abell
2255}, we aim to unveil their 3-dimensional location within the
cluster.}{We performed WSRT observations of \object{A2255} at 18, 21, 25, 85,
and 200~cm. The polarization images of the cluster were processed
through rotation measure (RM) synthesis, producing three final
RM cubes.}{The radio galaxies and the filaments at the
edges of the halo are detected in the high-frequency RM cube, obtained by combining the
data at 18, 21, and 25~cm. Their Faraday spectra show different
levels of complexity. The radio galaxies lying near by the cluster
center have Faraday spectra with multiple peaks, while those at large
distances show only one peak, as do the filaments. Similar RM
distributions are observed for the external radio galaxies and for the
filaments, with much lower average RM values and RM variance than
those found in previous works for the central radio galaxies.\\ The
85~cm RM cube is dominated by the Galactic foreground emission, but it
also shows features associated with the cluster. At 2~m, no polarized
emission from \object{A2255} nor our Galaxy is detected.}{The radial trend
observed in the RM distributions of the radio galaxies and in the
complexity of their Faraday spectra favors the interpretation that the
external Faraday screen for all the sources in \object{A2255} is the ICM. Its
differential contribution depends on the amount of medium that the
radio signal crosses along the line of sight. The filaments should
therefore be located at the periphery of the cluster, and their
apparent central location comes from projection effects. Their high
fractional polarization and morphology suggest that they are relics
rather than part of a genuine radio halo. Their inferred large
distance from the cluster center and their geometry could argue for an
association with large-scale structure (LSS) shocks.}


\authorrunning{Pizzo et al.}
\titlerunning{Deep RM tomography of the galaxy cluster \object{A2255}}
\maketitle
%          
%________________________________________________________________

\section{Introduction}

\indent Clusters of galaxies contain a conspicuous amount of gas
classified as intracluster medium (ICM). The dominant baryonic
component of the ICM is represented by the X-ray emitting thermal
plasma, amounting to about 15\%--17\% of the total gravitational mass
of the cluster \citep{2005xrrc.procE8.02F}. Additional nonthermal
components of the ICM, not obviously associated with individual radio
galaxies, have been detected in several clusters. \citet{fergiov}
classify them as {\it radio halos}: centrally located,~Mpc-scale, low
surface brightness, steep-spectrum radio sources with a regular
morphology, {\it radio relics}: elongated, steep-spectrum radio
sources often found at the periphery of clusters, and {\it radio
mini-halos}: diffuse, moderately large ($\sim500$~kpc) radio sources
surrounding powerful central dominant (cD) radio galaxies in cooling
core clusters. The existence of these extended nonthermal features in
galaxy clusters prove there are large-scale magnetic fields in
them. Their study is therefore the key to any comprehensive description
of the ICM.

Because they are the result of the synchrotron process, halos and relics are generally intrinsically polarized, assuming ordered
magnetic fields. While radio relics are indeed highly polarized
(20--40\%), radio halos generally do not show any significant
polarization \citep{fergiov}. This is thought to come from internal
depolarization along the line of sight and/or beam depolarization,
given the low resolution needed for detecting these
sources. The study of the polarization associated with halos and
relics is a powerful diagnostic tool that can be used to constrain the
strength and geometry of magnetic fields in clusters.

Important complementary information on cluster magnetic fields along
the line of sight can also be derived through the study of the
rotation measure distributions observed towards background and cluster
radio sources, since their polarized emission experiences Faraday
rotation while crossing the magnetized ICM. These studies have determined
values for cluster magnetic fields of a few~$\mu$G \citep[for a review
see ][and references therein]{carilli}.  In addition, stronger
magnetic fields are detected in the innermost regions of cooling core
clusters, where RM values of several hundreds of
rad~m$^{-2}$ \citep{2001ApJ...547L.111C} up to a few thousand
rad~m$^{-2}$ \citep{vogt} have been found.\\ An important technique
 recently developed to analyze and interpret
polarization data is RM-synthesis \citep{br2005}. By separating the
polarized emission as a function of Faraday depth, this technique can
give important information on the 3-dimensional structure of clusters
of galaxies. Such investigations become progressively more difficult
to carry out at low frequencies, where instrumental (i.e. beam and
bandwidth depolarization) and astrophysical effects (depolarization
occurring inside and outside the radio source) may reduce the observed
polarized emission.

\object{A2255} is a nearby \citep[z=0.0806,][]{struble} rich cluster, which has
been studied at several wavelengths. {\it ROSAT} X-ray observations
indicate that \object{A2255} is a non-cooling core cluster that has recently
undergone a merger
\citep{burns1995,1998ApJ...492...57D,fer,miller,sakelliou}. Optical
studies of \object{A2255} reveal the presence of kinematical substructures in
the form of several associated groups
\citep{2003ApJS..149...53Y}. This result, together with the high
ratio of velocity dispersion to X-ray temperature \citep[6.3
keV;][]{horner}, indicates a non-relaxed system. \\ At radio
wavelengths, \object{A2255} hosts a diffuse radio halo, a relic source, and
seven extended head-tail radio galaxies. On the basis of their
morphological properties at low frequency, they are named Goldfish,
Double, Original TRG, Sidekick, Bean, Beaver, and Embryo
\citep{har}. The former 4 radio galaxies lie near the cluster center,
while the others are located at a large projected distance: the Embryo
and the Beaver lie at 1.5~Mpc; the Bean at $\sim$ 3.5~Mpc. Most of
these radio galaxies have a narrow angle tail \citep[NAT,
][]{1976ApJ...203L.107R} morphology, which suggests a strong
interaction between the plasma ejected by the parent galaxy and the
ICM. Moreover, their tails show a random orientation with respect to
the cluster center, suggesting that they are in random orbits inside
the cluster \citep{fer}.\\ Sensitive Westerbork Synthesis Radio
Telescope (WSRT) observations have recently shown that \object{A2255} is
surrounded by several low surface brightness extended
features, which are possibly associated with large-scale structure
(LSS) shocks \citep{2008A&A...481L..91P}. A study of the cluster
at 21~cm has revealed that the radio halo is dominated by 3 filaments,
which are strongly polarized \citep[$\sim$ 20--40\%, ][]{gov}. The
distribution of their polarization angles indicates that the magnetic
field is ordered on scales up to 400~kpc.
           
In this paper we present WSRT polarimetric observations of Abell 2255 at 18,
21, 25, 85, and 200~cm. By extending the analysis presented in
\citet{2009ASPC..407..241P}, we study the polarimetric properties of the
radio galaxies and investigate the highly polarized emission of the halo,
unexpected for such a structure. RM-synthesis is used for analyzing the
data. The total intensity results from these observations were used to
investigate the spectral index properties of the structures belonging to the
cluster \citep{pizzospectrum}.

The outline of the paper is as
follows. Section \ref{observationsanddatareduction} describes
the main steps in the data reduction, the polarization calibration,
and the correction for the time-variable ionospheric Faraday
rotation. In Sect.~\ref{results} we present the RM cubes and
discuss the instrumental artifacts and the real signal in them. In
Sect.~\ref{theradiogalaxiesandthefilaments} we analyze the
results, determining the RM distributions of the radio galaxies and the
filaments and describing the characteristics of their Faraday
spectra. In Sect.~\ref{discussion} we discuss the
implications of our results on the nature of the radio filaments and
 summarize our work in Sect.~\ref{summaryandconclusions}.

In this paper we assume a $\Lambda$CDM~~ cosmology with $H_{0}=$
71~km~s$^{-1}$~Mpc$^{-1}$, $\Omega_{m}=0.3$, and
$\Omega_{\Lambda}=0.7$. All the positions are given in J2000. At the
distance of \object{A2255} ($\sim350$~Mpc), 1\hbox{$^{\prime}$} corresponds to
90~kpc.

%__________________________________________________________________

\begin{table*}
\caption{Observations overview.}
\label{observationsdetails}
\smallskip
\begin{center}
{\small
\begin{tabular}{cccccccc}
\hline
\hline
\noalign{\smallskip}
Wavelength & Frequency range &   \multicolumn{2}{c}{Calibrators}  & Observation ID & RT9--A     &Start date (UTC) & End date (UTC)\\
  (cm)     &    (MHz)        &   polarized  & unpolarized         &                &  (m)        &              &               \\
\noalign{\smallskip}
\hline
\noalign{\smallskip}
           &                &               &                     &                &             &                                          \\
18         & 1650--1795     &    3C286      &  CTD93              &  10602977      &  36         & 2006/06/22 16:43:20 & 2006/06/23 04:42:20\\
           &                &    3C138      &  3C48               &                &             &                                          \\
\noalign{\smallskip}
\hline
\noalign{\smallskip}
           &                &               &                     &                &             &                                          \\
21         & 1310--1470     &    3C286      &  CTD93              & 10602938       &  36         & 2006/06/21 16:47:10 & 2006/06/22 04:46:10\\
           &                &    3C138      &  3C48               &                &             &                                          \\ 
\noalign{\smallskip}
\hline
\noalign{\smallskip}
           &                &               &                     &                &             &                                          \\   
21         & 1310--1470     &    3C286      &  CTD93              & 10603082       & 36          & 2006/06/25 16:14:20 & 2006/06/26 04:06:00\\
(30 sources)&               &    3C138      &  3C48               &                &             &                                           \\
\noalign{\smallskip}
\hline
\noalign{\smallskip}   
           &                &               &                     & 10702691       & 36          & 2007/06/18 17:00:15 & 2007/06/19 04:58:50\\
           &                &    3C286      &  CTD93              & 10702898       & 54          & 2007/06/28 16:20:55 & 2007/06/29 04:19:25\\
25         & 1159--1298     &    3C138      &  3C48               & 10702958       & 72          & 2007/07/03 16:01:15 & 2007/07/04 03:59:45\\
           &                &               &                     & 10703066       & 90          & 2007/07/10 15:33:45 & 2007/07/11 03:10:35\\
\noalign{\smallskip}
\hline
\noalign{\smallskip}
           &                &               &                     &  10602227      & 36          & 2006/05/09 19:36:20 & 2006/05/10 07:35:20\\
           &                &               &                     &  10602239      & 48          & 2006/05/10 19:32:20 & 2006/05/10 07:31:20\\
           &                &   DA240       &  3C295              &  10602259      & 60          & 2006/05/11 19:28:20 & 2006/05/12 07:27:20\\
85         & 310--380       &   PSR1937+21  &  3C48               &  10602224      & 72          & 2006/05/16 19:08:40 & 2006/05/17 07:07:40\\
           &                &               &                     &  10602372      & 84          & 2006/05/19 18:57:00 & 2006/05/20 06:56:00\\
           &                &               &                     &  10602340      & 96          & 2006/05/22 18:45:10 & 2006/05/23 06:44:10\\
\noalign{\smallskip}
\hline
\noalign{\smallskip}
           &                &               &                     & 10702727      & 36          & 2007/06/21 16:48:15 & 2007/06/22 04:47:05\\
           &                &               &                     & 10703226      & 48          & 2007/07/19 14:58:05 & 2007/07/20 02:56:55\\
           &                &   DA240       &  3C295              & 10703186      & 60          & 2007/07/17 15:05:55 & 2007/07/18 03:04:45\\
200        & 115--175       &   PSR1937+21  &  3C48               & 10703004      & 72          & 2007/07/05 15:53:15 & 2007/07/06 03:52:05\\
           &                &               &                     & 10703056      & 84          & 2007/07/09 15:37:25 & 2007/07/10 03:36:15\\
           &                &               &                     & 10703099      & 96          & 2007/07/12 15:25:35 & 2007/07/13 03:24:25\\
\noalign{\smallskip}
\hline
\end{tabular}
}
\end{center}
\end{table*}

\begin{table}[h!]
\caption{Details of the flux calibrators.}
\label{calibrators}
\smallskip
\begin{center}
{\small
\begin{tabular}{ccccc}
\hline
\hline
\noalign{\smallskip}
Wavelength & ${\nu_0} ^a$ &       name         &   Flux   & $\alpha$\\
   (cm)    &   (MHz)      &                    &   (Jy)   &         \\     
\noalign{\smallskip}
\hline
\noalign{\smallskip}
18         &  1722   &           CTD93         &   4.4    &  --0.5 \\
\noalign{\smallskip}
\hline
\noalign{\smallskip}
21         &  1380   &           CTD93         &   4.9    &  --0.5 \\
\noalign{\smallskip}
\hline
\noalign{\smallskip}
25         &  1220   &           CTD93         &   5.2    &  --0.5  \\
\noalign{\smallskip}
\hline
\noalign{\smallskip}
85         &  346    &           3C295         &    63    &  --0.6  \\
\noalign{\smallskip}
\hline
\noalign{\smallskip}
200        &  148    &           3C295         &    95    &  --0.6  \\
\noalign{\smallskip}
\hline
\noalign{\smallskip}
\multicolumn{5}{l}{$^a$ central frequency of the observing bands.}
\end{tabular}
}
\end{center}
\end{table}

\section{Observations and data reduction}
\label{observationsanddatareduction}

The observations were conducted with the WSRT.  The array consists of fourteen
25~m~dishes on an east-west baseline and uses earth rotation to fully
synthesize the uv plane.  Ten telescopes (labeled from 0 to 9) are on fixed
mountings, 144~m apart; the four remaining dishes (labeled A, B, C, and D) are
movable along two rail tracks.\\  Our observations were performed using both the maxi-short
configuration and the standard configurations \citep{morgantiwsrt}. The first provides optimum imaging performance for very extended sources within a single track observation by relocating the movable telescopes such that a 36 m, 54 m, 72 m, and 90 m spacing are all observed. The standard configurations provide high dynamic-range imaging when combining the data obtained with multiple track observations. For
these we kept the distances between the movable telescopes constant (RTA--RTB
and RTC--RTD at 72~m and RTA--RTC at $9 \times 144$~=~1296 m), while changing the
distance RT9--RTA from 36 to 90 or 96~m~in fixed intervals. The baselines
therefore range from 36~m~to 2.7~km.\\ Table
\ref{observationsdetails} summarizes the observational
parameters. The observations were bracketed by two pairs of calibrators, one
polarized and one unpolarized, observed for 30 minutes each. The pointing
center of the telescope, as well as the phase center of the array was directed
towards RA~=~$17 ^{\rm h}13^{\rm m} 00^{\rm s}$, Dec~=~$+64\hbox{$^{\circ}$
}07\hbox{$^{\prime}$ }59\hbox{$\arcsec$}$, which is the radio center of
\object{A2255}.\\ The time sampling of the data is 30 s for the observations at 18~cm,
21~cm, 25~cm, and 85~cm, and 10~s~at 2~m. This is generally sufficient to
sample the phase fluctuations of the ionosphere and to avoid time smearing for
sources at the outer edge of the field. \\ The technical details about each
individual dataset will be discussed in the next sections. Here, we give a
short overview about the main steps taken in the data reduction. The data
were processed with the WSRT-tailored NEWSTAR reduction package
\citep{newstar} following the standard procedure: automatic interference
flagging, self-calibration, fast Fourier transform imaging, and CLEAN
deconvolution \citep{1974A&AS...15..417H}. Further flagging based on the
residual data after self-calibration and model subtraction was done after each
self-calibration iteration.  An on line Hamming taper was used to lower the
distant spectral side-lobe level \citep{1978ieee...66...51H}. This makes the two
contiguous channels highly dependent, so the final analysis was done
using only the odd channels.\\ In Table \ref{calibrators} we report
the fluxes adopted for the calibrators at the different observing
wavelengths. We computed them across the entire bands by assuming a spectral
index of $\alpha~=~-0.5$ for \object{CTD93} and $\alpha~=~-0.6$ for \object{3C295}. At 18~cm, 21~cm, 25~cm, and 85~cm the fluxes are based on the
\citet{1977A&A....61...99B} scale and have an uncertainty of about 1\%. At 2~m
the fluxes have been extrapolated, so they are believed to be
accurate at about 5\%. The details on the polarization calibration are
discussed in Sect.~\ref{bandpassandpolarisationcalibration}. \\
%Because of its equatorial mount -no change in parallactic angle- the WSRT
%suffices of an observation of one polarised and one unpolarised calibrator in
%order to calibrate the leakages. Hoewever, our observations were bracketed
%by two pairs of calibrators, one polarised and one unpolarised, observed for 30
%$minutes each.

\subsection{18~cm and 21~cm datasets}

At 18~cm and 21~cm, \object{A2255} was observed for a single 12$^{\rm h}$ run
with the ``maxi-short'' configuration of the WSRT. To distinguish between the polarized cluster emission and
Galactic emission, 35 background sources were also observed at 21
cm. They were selected on the basis of their polarized ($>$ 1.2~mJy) NVSS
polarization fluxes and within an area of about 4 degrees
diameter centered on the cluster.

At these wavelengths, the feeds are sensitive to frequencies between 1310 and
1795~MHz. The backend covers this frequency range with 8 tunable, 20~MHz wide
bands. At 18~cm the bands are centered at 1659, 1677, 1695, 1713, 1731, 1749,
1767, 1785~MHz and at 21~cm at 1459, 1437, 1419, 1401, 1379, 1359, 1339, and 1320
MHz. The bands have a 2~MHz overlap to provide a continuous frequency
coverage. Each sub-band is covered by 64 channels in 4 cross-correlations to
recover all the Stokes parameters.

Due to strong RFI during the observations of 5 of the 35 sources
surrounding \object{A2255}, the final analysis was restricted to 30
sources.  For each of the 3 datasets, the total amount of flagged data
was $\sim10$\%.  At 18~cm the resolution is $
11\hbox{\arcsec} \times 12\hbox{\arcsec}$ and at 21~cm $ 12\hbox{\arcsec}
\times 13\hbox{\arcsec}$.

\subsection{25~cm dataset}

At 25~cm, \object{A2255} was observed for $4 \times 12^{\rm h}$. By stepping
the 4 movable telescopes at 18~m increments from 36 to 90~m, the first
grating lobe due to the regular 18~m~baseline increment is pushed to
a radius of $\sim1^{\circ}$. This allows faithful
reconstruction of large-scale cluster-wide emission.\\
At this wavelength, the receiving band is divided into 8 contiguous, 
slightly overlapping sub-bands of 20~MHz centered at 1169, 1186, 1203,
1220, 1237, 1254, 1271, and 1288~MHz. Each sub-band is covered by 64
channels in 4 cross-correlations to recover all the Stokes parameters.\\
Because of limitations in the software when handling files larger than
2.15 GB, the data was reduced for each frequency band
independently.  Thus, the original dataset was processed per sub-band,
but each containing the $4 \times 12^{\rm h}$ of observation.
Because of strong RFI, 3 out of the original 8 frequency sub-bands
were not used for the polarization imaging. The total amount of
unused or flagged data was $\sim40$\%. At 25~cm, the resolution is
$14\hbox{\arcsec} \times 15\hbox{\arcsec}$.

\subsection{85~cm dataset}

At 85~cm, \object{A2255} was observed for $6 \times 12^{\rm h}$. By combining the data of $12^{\rm h}$ runs performed with particular WSRT configurations, the first grating lobe in the final maps is pushed away to beyond the 10 dB point of the primary beam. This ensures that the diffuse emission that fills the primary beam is not `self-confused'. We
used 6 array configurations, in which the four movable telescopes were
stepped at 12~m increments (i.e. half the dish diameter). The shortest
spacing was stepped from 36 to 96~m, providing continuous
uv coverage with baselines ranging from 36 to 2760~m.\\
The receiving band covers the frequency range 310--380~MHz and is
divided into 8 sub-bands of 10~MHz centered at 315, 324, 332, 341, 350,
359, 367, and 376~MHz. Using recirculation in the backend, the number
of frequency channels was increased to 128 channels in 4
cross-correlations to recover all the Stokes parameters. The initial
dataset was split into 16 smaller datasets, each containing $6\times12^{\rm h}$, but
only half a sub-band. Approximately 25\% of the data were flagged. The
resolution at 85~cm is $54\hbox{\arcsec} \times 64\hbox{\arcsec}$.

 \subsection{2~m dataset}
\label{2mdataset}

At 150~MHz, \object{A2255} was observed for $6 \times 12^{\rm h}$. During each session, the polarized calibrator \object{PSR1937+21} was
observed for seven times to monitor and correct for the ionospheric Faraday
rotation (see Sect.~\ref{ionosphericfaradayrotation}).

The LFFE (low-frequency front end) at the WSRT is equipped with
receivers sensitive in the frequency range 115--175~MHz. The full band
is covered by eight sub-bands of 2.5~MHz centered at 116, 121,
129, 139, 141, 146, 156, and 162~MHz. Using a factor of eight
recirculations for these observations, each frequency band was divided
into 512 channels. 
Given the large size of the final dataset ($\sim250$~GB), the
observations have been split in 240 smaller datasets, each containing
one fifth of a sub-band for one spacing.
Due to serious correlator problems during the night of 17--18
July 2007 (RT9--RTA~=~60 m), the final data reduction was done using
$5 \times$ 12$^{\rm h}$ of observation. The total amount of flagged data
was $\sim35$\%. The resolution at 150~MHz is $163\hbox{\arcsec}
\times 181\hbox{\arcsec}$.

\begin{figure*}[!]
\begin{center}
\includegraphics[scale=.60]{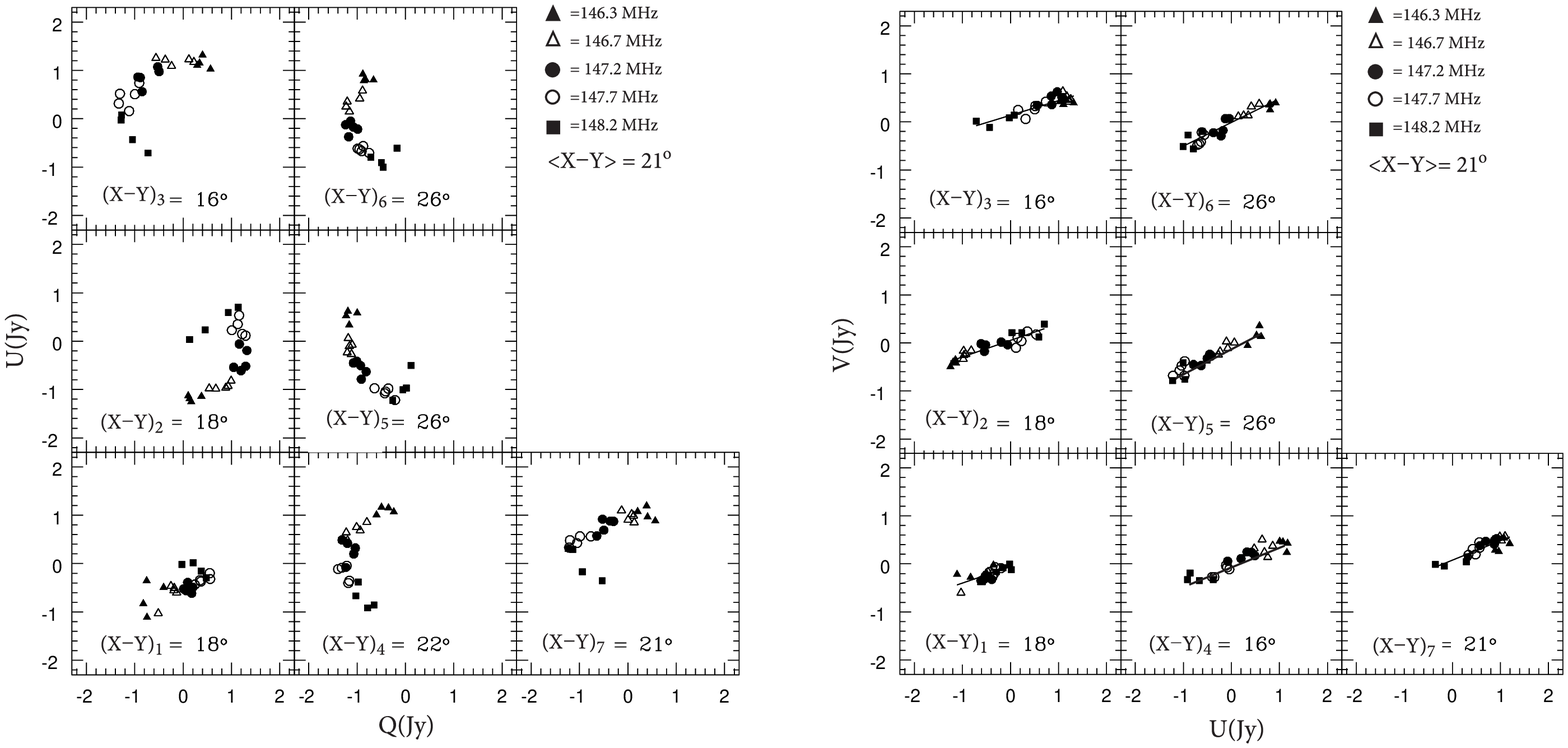}
\includegraphics[scale=.60]{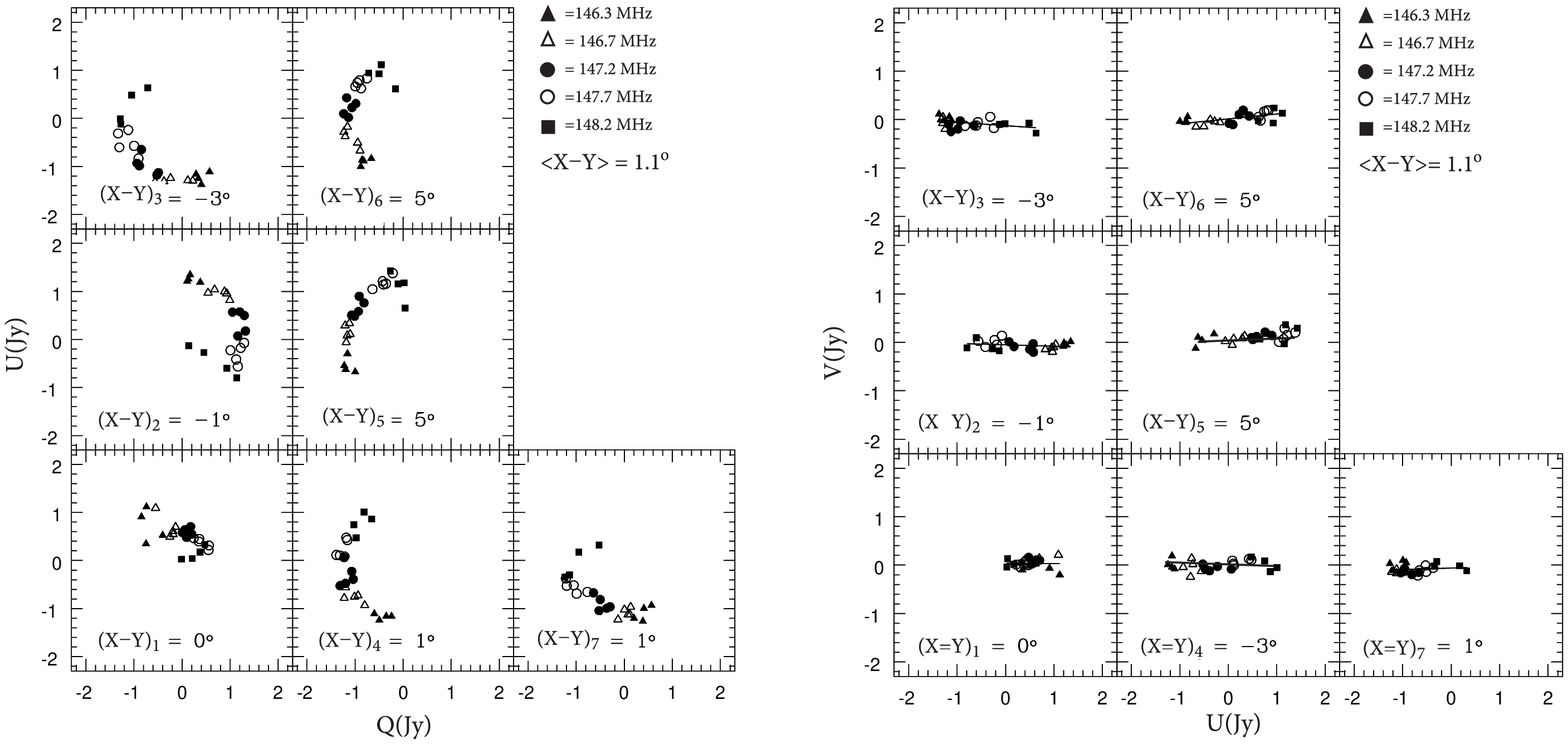}
\caption{Examples of the QU and UV plots for the 7 short observations of
  PSR1937 during the 12$^{\rm h}$ session of June 22, 2007 (spacing
  configuration: RT9--RTA~=~36 m) before (top panel) and after (bottom panel) the
  leakage correction. The plots refer to the frequency band 146--148~MHz. The
  symbols identify the different sub datasets in which the original dataset
  has been split (see text). Each point has a value averaged over a group of
  10 frequency channels. Since \object{PSR1937+21} has an RM $\sim+8.19$~rad~m$^{-2}$,
  the polarization vector in the QU plane rotates about 80\hbox{$^{\circ}$ }
  within this frequency range.}
\label{qupoints}
\end{center}
\end{figure*}

\label{ionospherepage}

\subsection{Polarization calibration}
\label{bandpassandpolarisationcalibration}

The WSRT telescopes are equipped with a pair of orthogonal feeds
(XY). All four cross correlations between the incident signals are
formed.\\
The overall (on-axis) instrumental leakages at the WSRT, typically 1\%--2\%,
were calibrated using an unpolarized calibrator.  The instrumental
polarization corrections were transferred to the polarized calibrator, which
was then used to phase-align the two orthogonal linear polarizations.
Finally, the polarization corrections were transferred to the target
source. The beam pattern of the WSRT has strong instrumental off-axis polarization \citep[e.g. ][]{2005A&A...441..931D,2008A&A...479..903P}, which changes with frequency with a dominant 17~MHz pattern. This instrumental polarization might be related to standing waves between the dish and the front end. We did not correct for this effect, as it requires software that is not yet available.
 The polarization calibration followed the standard steps described
in \citet{1996A&AS..117..137H} and \citet{1996A&AS..117..149S}. Most of the
calibration procedure was carried out by scripts calling NEWSTAR routines, but
manual corrections had to be applied at 85~cm and 2~m because the scripts
failed to find a physical solution.

The polarized signal is described by the polarization vector {\it p},
whose intensity (P) and angle ($\chi$) are given by
\begin{equation}
 P = \sqrt{Q^2 + U^2 + V^2}\;, 
\end{equation}
\begin{eqnarray}
 \chi = \frac{1}{2} {\rm tan}^{-1} \frac{\rm U}{\rm Q} \;, 
\end{eqnarray}
where Q, U, and V are the Stokes parameters. When calibrating the data
of the polarized calibrator, one must verify that {\it p} rotates in
the QU plane according to the known RM value of the polarized
calibrator. The amount of rotation $\Delta \Phi$ is given by
\begin{equation}
\Delta \Phi~{\rm [rad]} = {\rm RM} \times \lambda_{\rm [m]} ^2\;\;\;\;\;\;\;~~,
\end{equation}
where $\lambda$ is the observing wavelength. No signal in V is expected to be
detected because most of the astrophysical sources only shows linear polarization.\\ A phase difference between the X and Y channels, however, will rotate
U into V. The X--Y phase difference of the polarization calibrator is given by
\begin{equation}
\label{x-y}
\delta_{X-Y} = {\rm tan}^{-1} \frac{\rm V}{\rm U} + n\pi~~.
\end{equation}
Equation \ref{x-y} has two solutions: one where $n$ is even
and one where $n$ is odd.  The wrong solution flips the sign of the RM
of the calibrator; therefore, if the sign is known {\it a priori}, it
is trivial to select the correct solution.\\ The polarization
calibration was performed at 85~cm using DA240 as polarized
calibrator. Since this source has an ${\rm RM} =
+3.33\pm0.14$~rad~m$^{-2}$ \citep{2008A&A...489...69B}, we verified
that at the end of the calibration $p$ rotates $\sim1$~rad from 381
MHz to 315~MHz and that there is no residual V signal left.\\ At
150~MHz, we used \object{PSR1937+21} as polarized calibrator. The source was
observed 7 times during each observing session, in order to monitor
its RM changes. This pulsar has an RM of
+8.19~$\pm$~0.08~rad~m$^{-2}$ (see
Sect.~\ref{ionosphericfaradayrotation}) and a polarized
intensity of $\sim1.5$~Jy at 146~MHz.\\ The results of the procedure
are shown in Fig.~\ref{qupoints}, where we present the
situation in one band (146 MHz) for the first observing night
(RT9--RTA~=~36 m) before (top panel) and after (bottom panel) the
polarization calibration. To increase the signal to noise ratio and
have a better determination of the measured Q, U, and V fluxes, the
data were averaged every 10 channels. Before the correction, the QU
points lie on a circle with approximately the correct radius, but with
the wrong order: going from lower frequencies to higher frequencies, 
the circle should be described clockwise, while here we see the
opposite situation. Consequently, in the UV plots there are points
that have a non-zero V value. The signal in V should be transferred to
U. The X--Y phase difference, calculated using a linear fit for the UV
points, defines the correction that must be applied to the data. This
is determined for each time cut and averaged during the night. After
the correction, all the points have approximately a zero V value and
describe a circle in the QU plane with the correct rotation direction.

The signal was also corrected for the ionospheric Faraday rotation,
which is relevant at low frequencies. This is described below.

\subsection{ Ionospheric Faraday rotation }
\label{ionosphericfaradayrotation}

The Earth's ionosphere has density fluctuations that mostly affect low-frequency observations, introducing a direction-dependent variation in
the phase of the signal received by the interferometer. In addition,
it also affects the polarization, giving rise to time-variable Faraday
rotation. This often amounts to several
turns of the polarization vector in the QU plane at low
frequency. This effect can significantly depolarize the signal. The
ionospheric Faraday rotation at the WSRT is typically a few tenths of
rad~m$^{-2}$ during night time. Daytime values of 5~rad~m$^{-2}$ are
also possible during the solar maximum \citep{2008A&A...489...69B}. A
variation of just 0.5~rad~m$^{-2}$ in ionospheric Faraday depth
corresponds to a change in polarization angle of the signal of
$\sim200^{\circ}$ at 115~MHz, the lowest frequency of the 2~m
observations. The correction for this effect is important, because it
could significantly affect the results of the polarization imaging.

\begin{figure*}
\begin{center}
\includegraphics[scale=.40]{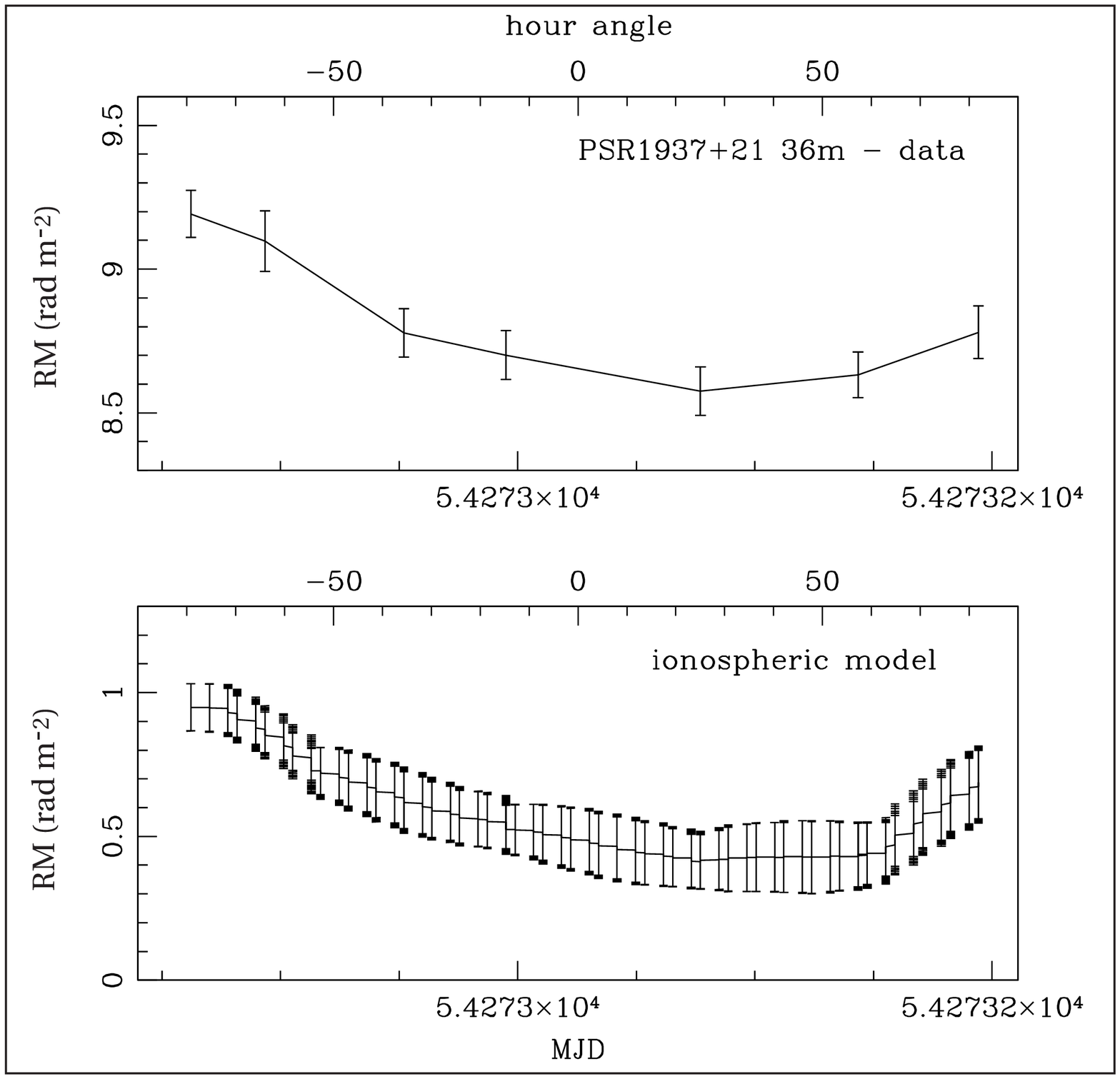}
\includegraphics[scale=.40]{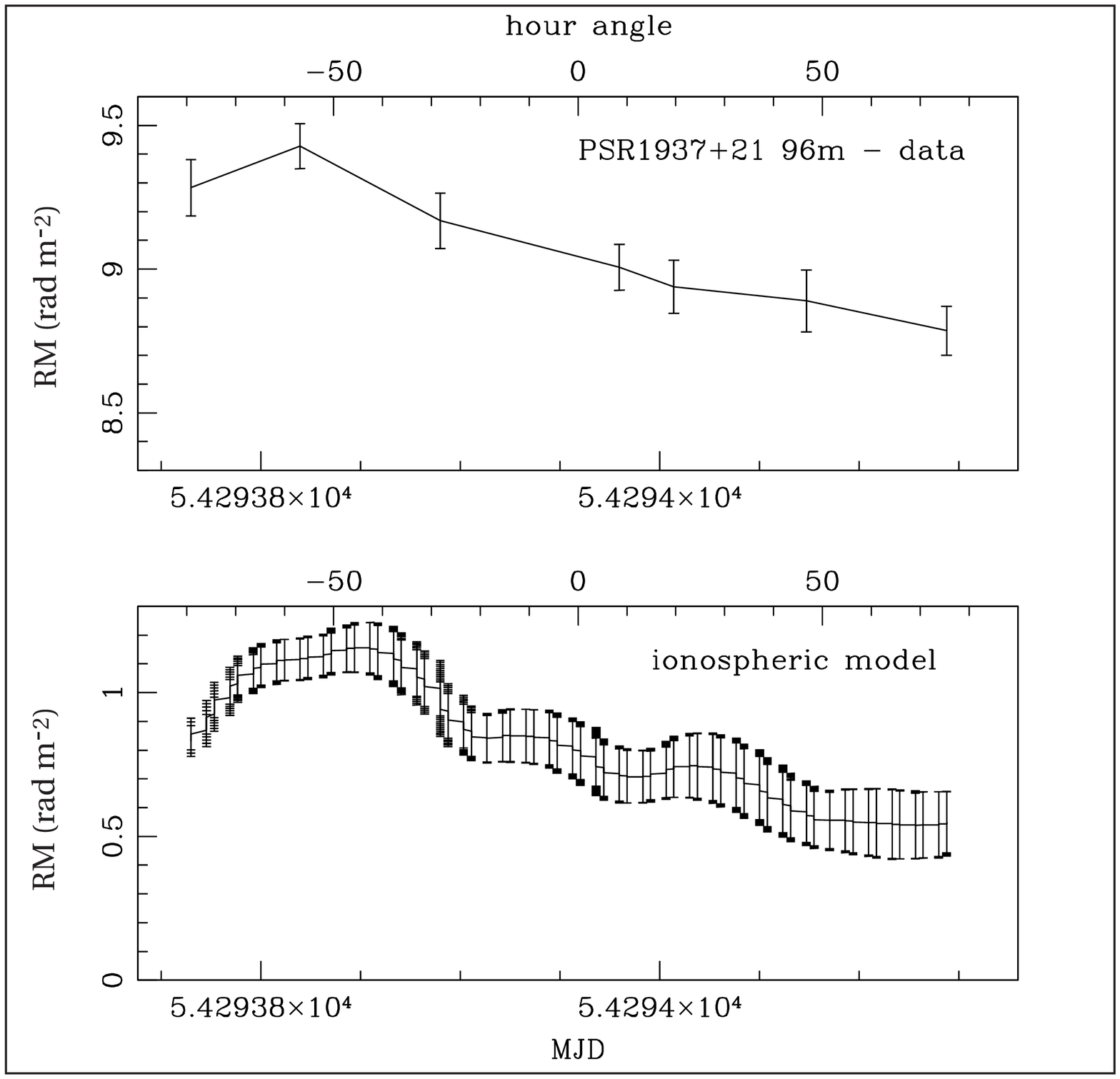}
\caption{Comparison between the data and the model predictions for the RM
  variations of \object{PSR1937+21} for the nights when the WSRT was observing
  in the 36~m~(left panel) and 96~m (right~ panel) configurations. Along the
  horizontal axes, the time of the observation is expressed in modified Julian
  days (MJD) and hour angle (HA), going from $-90^{\circ}$ to
  $+90^{\circ}$. Along the vertical axis we plot the rotation measure, which
  is the sum of the contributions of the pulsar plus the ionosphere for the
  data, while it is due to the only ionosphere in the model.}
\label{corrections1937+21}
\end{center}
\end{figure*}

The correction for the ionospheric Faraday rotation was only applied to the
2~m dataset, which is the one most affected by the problem. It was computed using
global GPS total ionospheric electron content (TEC) data and an analytical
model for the geomagnetic field. The GPS-TEC data were provided by the Center
for Orbit Determination in Europe (CODE) of the Astronomical Institute of the
university of Bern, Switzerland. The geomagnetic field was computed using the
International Geomagnetic Reference Field (IGRF), which consists of a series
of mathematical models of the Earth's main field and of its annual rate of
change. The ionosphere was modeled as a spherical shell with a finite
thickness and uniform density at an altitude of 350~km above the mean sea
level. The validity of the model was first checked on the polarized
calibrator. We expect to see agreement between the computed and the
observed RM variations towards \object{PSR1937+21}, monitored 7 times each night
(8 times during the session with configuration RT9--RTA~=~84~m). In
Fig.~\ref{corrections1937+21} we show the comparison between the
predictions and the data for two of the six observing nights. The
  expected uncertainty in the RM towards the pulsar, based on the TEC RMS
  uncertainty published by CODE, is typically 0.07--0.1~rad~m$^{-2}$ for each
  data point during the observation. For each cut,
the observed RM of the pulsar agrees with the predictions within 5\%. The only
major differences are visible for the first cut during the nights when the
WSRT was observing with the 84 and 96~m configurations. 

The total Faraday depth of a source as a function of time
($\phi_{tot}(t)$) stems from the time-dependent ionospheric Faraday
rotation ($\phi_{ion}(t)$) plus the constant contribution of the
observed source ($\phi_{src}$):
\begin{equation}
\label{phi}
\phi_{tot}(t)~=~\phi_{ion}(t) + \phi_{src} \; .
\end{equation}
Through Eq.~\ref{phi} we calculate that \object{PSR1937+21} has an ${\rm RM}~=~+8.19
\pm 0.08$~rad~m$^{-2}$, which represents the most accurate determination of
the RM of this pulsar to date.  From previous observations at 2~m, where no
correction for the ionospheric Faraday rotation was applied, A.G. de Bruyn
found a rotation measure for the pulsar of $+8.5\pm0.5$~rad~m$^{-2}$.  At
350~MHz and after correcting for the rotation due to the ionosphere,
\citet{Brentjens} derived a value of ${\rm RM} =+7.86\pm0.20$~rad~m$^{-2}$.

After checking the validity of the model for the polarized calibrator,
we computed the expected ionospheric RM variations towards the center
of \object{A2255}, which is located in a different area of the sky, and we
removed the ionospheric contribution from the data.

\subsection{ Polarization imaging and RM cubes }
\label{imagingQU}

After performing the polarization calibration procedure described in
Sects. \ref{bandpassandpolarisationcalibration}--\ref{ionosphericfaradayrotation},
we produced the Q and U images of the target source.
For the 30 background sources observed at 21~cm, we determined the
polarization angle in each sub-band and plotted it as a function of
$\lambda^2$, obtaining the values of their RM through a linear
fit.
For the cluster observations, we made Q and U image cubes and
inspected them to remove the channel maps affected by residual RFI
and/or problems related to specific telescopes. For the 18~cm, 21~cm,
and 25 cm datasets, a total of 193, 217, and 133 channel images were
used for further analysis. At 85~cm we flagged 48 channels from the
available 448. At 2~m we used 1523 images out of the available
1904. Table \ref{noisesQU} shows the average noise level
($\sigma$) for each Q and U channel for each dataset. This was estimated
by subtracting adjacent channel maps. The final images were carried
through for further processing in RM-synthesis. To improve the
resolution in RM space, as well as the sensitivity, we decided to
combine the 18~cm, 21~cm, and 25~cm datasets, giving 543 channel
images as input to the RM-synthesis.

In the wide range of frequencies of each observation the primary beam
width changes significantly across the band\footnote{The WSRT primary
beam is approximated well by $G(\nu,r) = cos^6(c\nu r)$, where $c$ is
$\sim0.064$, $\nu$ the observing frequency in~MHz, and $r$ the
radius from the pointing center in degrees.}. The input images to
RM-synthesis were not corrected for the primary beam attenuation. By
doing this a constant spatial noise level across the map was retained.

\begin{table}
\caption{Noise levels of the Stokes I and Q-U channel maps at the different wavelengths of our observations.}
\label{noisesQU}
\smallskip
\begin{center}
{\small
\begin{tabular}{ccc}
\hline
\hline
\noalign{\smallskip}
Wavelength  & $\sigma_{I}$      &    $\sigma_{U-Q}$     \\
  (cm)      & (mJy beam$^{-1}$)   &  (mJy beam$^{-1}$)    \\
\noalign{\smallskip}
\hline
\noalign{\smallskip}
18          & 0.24                 &        0.22           \\
\noalign{\smallskip}
\hline
\noalign{\smallskip}
21          & 0.18                 &        0.15            \\
\noalign{\smallskip}
\hline
\noalign{\smallskip}
21          & 0.18                 &        0.15             \\
(30 sources)&                      &                         \\
\noalign{\smallskip}
\hline
\noalign{\smallskip}
25         & 0.17                  &        0.16              \\
\noalign{\smallskip}
\hline
\noalign{\smallskip}
85         & 1.6                   &         0.8              \\
\noalign{\smallskip}
\hline
\noalign{\smallskip}
200        &  78                   &           30   \\
\noalign{\smallskip}
\hline
\end{tabular}
}
\end{center}
\end{table}

\subsection{RM-synthesis}

After imaging, RM-synthesis was applied to the Q and U cubes. This
technique derotates the Q-U vectors for each pixel in each channel
image, in order to compensate for a certain assumed rotation
measure. After derotation, the channel images are
averaged. RM-synthesis maximizes the sensitivity to radiation at the
assumed Faraday depth, because that emission is added coherently. All
other emission will add up coherently only in part, hence the sensitivity
to emission not at the assumed Faraday depth is reduced. RM-synthesis
has already been mentioned by \citet{1966MNRAS.133...67B} and is known by the
pulsar community \citep{2003A&A...398..993M,2004ApJS..150..317W}, but
only recently it has been applied to large datasets. A full
description of how this technique works is presented by
\citet{br2005}, but in the following, we give a short overview.

After computing the complex polarization P~=~Q + iU, one multiplies it
by a complex phase factor to perform the rotation. \citet{br2005}
find that a general inversion of the polarization as a function of
wavelength squared is given by
\begin{equation}
\label{1}
\{F*R\}(\phi)=\frac{\int_{-\infty}^{+\infty} W(\lambda^2)P(\lambda^2) e^{-2i\phi (\lambda^2 - \lambda_0 ^2) d\lambda^2}}{\int_{-\infty}^{+\infty} W(\lambda^2) d\lambda} \;,
\end{equation}
where $\phi$ is the Faraday depth, defined as
\begin{equation}
\label{rmequation}
\phi ~~{\rm [rad~m]^{-2}} = 812 \int_{0}^{L_{\rm [kpc]}} n_{e {\rm ~[cm^{-3}]}} {\textbf B_{\parallel~[\mu{\rm G}]}} \cdot d{\textbf l}~~,
\end{equation}
where B$_{\parallel}$ is the magnetic field component along the line
of sight ($l$) and $n_e$ the electron density. In equation
\ref{1}, $F(\phi)$ is the emission as a function of Faraday
depth, $R(\phi)$ the rotation measure structure function (RMSF),
$\lambda_0$ the wavelength to which all vectors are derotated, and
$W(\lambda^2)$ the weight function and represents the sensitivity
as a function of wavelength squared.

If the width of a channel is much smaller than the width of the band,
the weight function can be approximated by a sum of $\delta$
functions, and equation \ref{1} becomes
\begin{equation}
\{ F * R \} (\phi_k)= \frac{\sum_{i=1}^{N} w_i P_i e^{-2i\phi(\lambda_i ^2 - \lambda_0 ^2)}}{\sum_{i=1}^{N}w_i}~~,
\end{equation}
where $P_i = P(\lambda_i ^2)$ and $w_i$ is the weight of a data
point. The RMSF is given by
\begin{equation}
R(\phi)= \frac {\sum_{i=1}^{N} w_i e^{-2i\phi (\lambda_i ^2 - \lambda_0 ^2)}}{\sum_{i=1}^{N}w_i}~~.
\end{equation}
Here, $\lambda_0$ is arbitrary and we chose the wavelength corresponding to
the weighted average $\lambda^2$.

\begin{figure}
\begin{center}
\includegraphics[scale=.9]{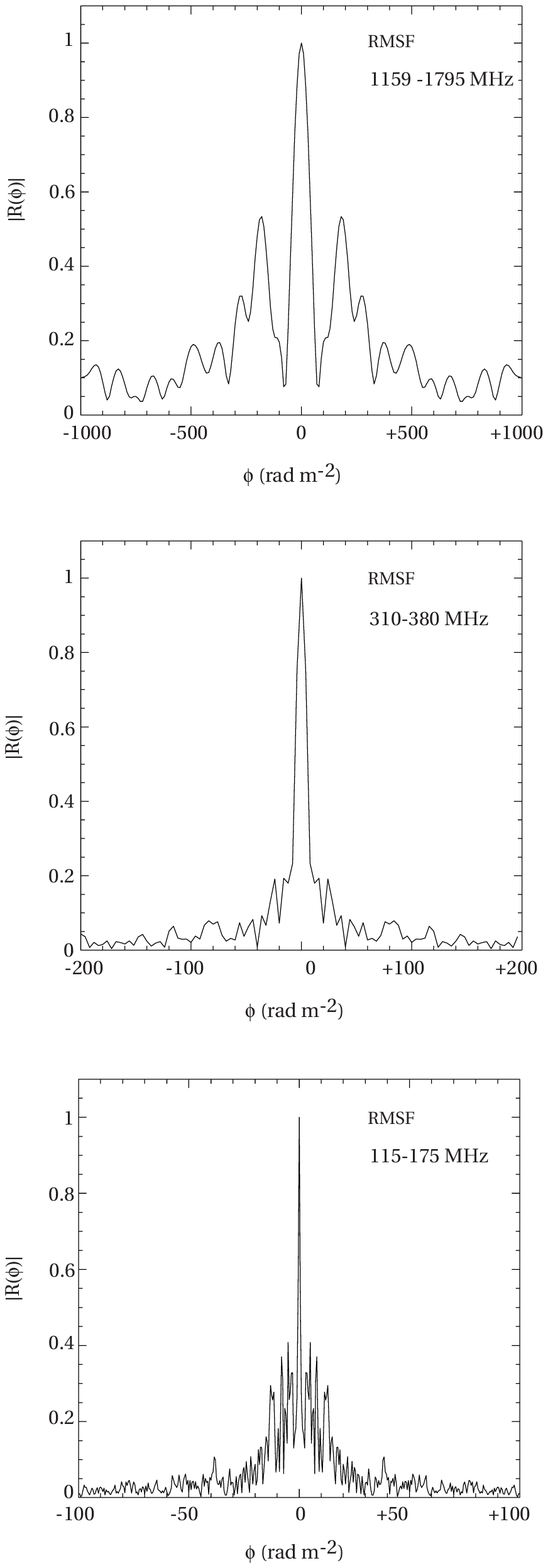}
\caption{Absolute value of the RMSF corresponding to the frequency
coverage of the observations: 18~cm +21~cm +25~cm dataset (top panel),
85~cm dataset (middle panel), and 2~m dataset (bottom panel).}
\label{RMSF}
\end{center}
\end{figure}

When planning a rotation-measure experiment, three main parameters are
involved, namely the channel width $\delta \lambda^2$, the width of the
$\lambda^2$ distribution $\Delta \lambda^2$, and the shortest wavelength
squared $\lambda_{min} ^2$. They respectively determine the maximum observable
Faraday depth, the resolution in Faraday space, and the largest scale in
Faraday space to which the observation is sensitive. If we assume a top hat weight
function that is 1 between $\lambda_{min} ^2$ and $\lambda_{max} ^2$ and zero
elsewhere, the estimates of the FWHM of the main peak of the RMSF, the scale
in Faraday space to which the sensitivity has dropped to 50\%, and the maximum
Faraday depth to which one has more than 50\% sensitivity are approximately
\begin{eqnarray}
\delta \phi \sim \frac{2 \sqrt{3}}{\Delta \lambda^2}~~,\nonumber 
\end{eqnarray}
\begin{eqnarray}
{\rm max scale} \sim \frac{\pi}{\lambda_{min} ^2}~~,\nonumber  
\end{eqnarray}
\begin{eqnarray} 
\mid \mid \phi_{\rm max} \mid \mid \sim \frac{\sqrt{3}}{\delta \lambda^2}~~. 
\end{eqnarray}
In Table \ref{FWHM}, we list the parameters for the three
ranges of wavelengths used for RM-synthesis.  In
Fig.~\ref{RMSF} we show the RMSF of the 18cm+21cm+25cm, 85~cm,
and 2~m datasets. \\ Due to the sparse sampling in $\lambda^2$, the RM cubes
have ~sidelobes of $\sim40$\%, $\sim15$\%--20\%, and
$\sim60$\% at 2~m, 85~cm, and 18cm+21cm+25cm, respectively. These
sidelobes can be cleaned through a 1-dimensional deconvolution,
which is an extension of \citet{1974A&AS...15..417H} CLEAN to the
complex domain. We used public available code\footnote {The code is
available at http://www.astron.nl/$\sim$heald/software.}
made available by George Heald to perform the deconvolution. The
procedure follows.
\begin{enumerate}
\item{An each spatial pixel, the complex ($Q(\phi),U(\phi)$) spectrum
is cross-correlated with the complex RMSF and the location of the peak
in Faraday space ($\phi_m$) is noted.}
\item{If P($\phi_m$) is greater than the cutoff defined by the user,
then a shifted and scaled version of the RMSF is subtracted from the
complex ($Q(\phi),U(\phi)$) spectrum.}
\item{The complex scaling factor is stored as ``clean component'';} 
\item{steps 1--3 are repeated until one reaches the cutoff value,
which we adopted as 1 $\sigma$.}
\item{Finally, the clean components are restored with a restoring
beam with a specified FWHM and added to the residual F($\phi$).}
\end{enumerate}

\begin{table}
\caption{Parameters of the three RM cubes.}
\label{FWHM}
\smallskip
\begin{center}
{\small
\begin{tabular}{cccc}
\hline
\hline
\noalign{\smallskip}
Wavelength range     & FWHM               &   max scale$^{~a}$       &   $\mid\mid\phi_{\rm max}\mid\mid$$^{~b}$  \\
    (m)              &~rad~m$^{-2}$       &  ~rad~m$^{-2}$           &   ~rad~m$^{-2}$                            \\
\noalign{\smallskip}
\hline
\noalign{\smallskip}
   0.17--0.26         &  $\sim$ 90         &    $\sim$ 109            &     $\sim$ 26000               \\   
\noalign{\smallskip}
\hline
\noalign{\smallskip}
   0.79--0.96         &  $\sim$ 12         &    $\sim$ 5              &     $\sim$ 3200                \\ 
\noalign{\smallskip}
\hline
\noalign{\smallskip}
   1.8--2.6           &  $\sim$ 1          &    $\sim$ 1              &     $\sim$ 2700                \\ 
\noalign{\smallskip}
\hline
\noalign{\smallskip}
\noalign{\smallskip}
\multicolumn{4}{l}{$^a$ Scale in Faraday space to which the sensitivity has dropped}\\
\multicolumn{4}{l}{~~~to 50\%.}\\
\multicolumn{4}{l}{$^b$ Maximum Faraday depth to which one has more than 50\%}\\
\multicolumn{4}{l}{~~~sensitivity.}
\end{tabular}
}
\end{center}
\end{table}
%%----------------------------------------------------------------------

\section{Results}
\label{results}

\begin{table*}
\caption{Parameters of the 30 sources observed at
21~cm within a radius of 2$^{\circ}$ from the center of the cluster.}
\label{30sourcestable}
\smallskip
\begin{center}
{\small
\begin{tabular}{ccccccc}
\hline
\hline
\noalign{\smallskip}
Source     &      RA     &     DEC     &     RM           & F$_i ^{~a}$             &   F$_p ^{~b}$               &  D$ ^{~c}$        \\
number     &    (J2000)  &   (J2000)   &   rad~m$^{-2}$   & mJy~beam$^{-1}$         &   mJy~beam$^{-1}$           &   (Mpc)      \\
\hline
\noalign{\smallskip}
   1       &    16$^{\rm h}55^{\rm m}45^{\rm s}$  &    64$^{\circ}44^{\prime}06{\arcsec}$  &   $42  \pm       9   $                    & 14.6   &   1.2   &   23.5\\
   2       &    16$^{\rm h}56^{\rm m}39^{\rm s}$  &    63$^{\circ}16^{\prime}59{\arcsec}$  &   $13  \pm        4  $                    & 13.6   &   1.41  &   22.5 \\
   3       &    16$^{\rm h}56^{\rm m}56^{\rm s}$  &    63$^{\circ}44^{\prime}26{\arcsec}$  &   $-15 \pm        5  $                    & 22.5   &   3.3   &   21.8 \\
   4A, 4B  &    16$^{\rm h}58^{\rm m}43^{\rm s}$  &    62$^{\circ}57^{\prime}16{\arcsec}$  &   $-3  \pm        6, -12 \pm        6  $  & 29.2, 8.8   &   2.3, 1.5   &   20.3  \\
%  4B      &    16$^{\rm h}58^{\rm m}43^{\rm s}$  &    62$^{\circ}57^{\prime}16{\arcsec}$  &   $-12 \pm        6  $                    &   \\
   5A, 5B  &    16$^{\rm h}58^{\rm m}47^{\rm s}$  &    62$^{\circ}55^{\prime}50{\arcsec}$  &   $2   \pm        6, 0   \pm        3  $  & 31.4, 8.0   &   2.6, 2.2   &   20.3   \\
%  5B      &    16$^{\rm h}58^{\rm m}47^{\rm s}$  &    62$^{\circ}55^{\prime}50{\arcsec}$  &   $0   \pm        3  $                    &    \\
   6A, 6B  &    17$^{\rm h}00^{\rm m}48^{\rm s}$  &    63$^{\circ}07^{\prime}43{\arcsec}$  &   $-4  \pm        3, -11 \pm        11 $  & 6.29, 5.46   &   2.3, 1.3  &   17.3    \\
%  6B      &    17$^{\rm h}00^{\rm m}48^{\rm s}$  &    63$^{\circ}07^{\prime}43{\arcsec}$  &   $-11 \pm        11 $                    &     \\
   7       &    17$^{\rm h}05^{\rm m}36^{\rm s}$  &    63$^{\circ}19^{\prime}58{\arcsec}$  &   $-7  \pm        11 $                    & 5.74   &   1.3   &   10.9   \\
   8       &    17$^{\rm h}02^{\rm m}33^{\rm s}$  &    62$^{\circ}45^{\prime}12{\arcsec}$  &   $11  \pm        4  $                    & 65.8   &   2.0   &   15.9  \\
   9       &    17$^{\rm h}06^{\rm m}38^{\rm s}$  &    62$^{\circ}39^{\prime}22{\arcsec}$  &   $12  \pm        4  $                    & 3.8    &   1.2   &   11.7  \\
   10      &    17$^{\rm h}08^{\rm m}50^{\rm s}$  &    62$^{\circ}27^{\prime}43{\arcsec}$  &   $30  \pm        6  $                    & 9.9    &   1.2   &   10.6  \\
   11      &    17$^{\rm h}09^{\rm m}01^{\rm s}$  &    62$^{\circ}25^{\prime}39{\arcsec}$  &   $12  \pm        10 $                    & 20.2   &   1.3   &   10.6  \\
   12      &    17$^{\rm h}12^{\rm m}38^{\rm s}$  &    62$^{\circ}18^{\prime}22{\arcsec}$  &   $54  \pm        16 $                    & 9.1    &   1.2   &   9.9  \\
   13      &    17$^{\rm h}23^{\rm m}14^{\rm s}$  &    62$^{\circ}41^{\prime}29{\arcsec}$  &   $30  \pm        5  $                    & 18.5   &   3.0   &   15.9  \\
   14      &    17$^{\rm h}24^{\rm m}02^{\rm s}$  &    63$^{\circ}06^{\prime}16{\arcsec}$  &   $47  \pm        11 $                    & 27.1   &   1.2   &   15.9  \\
   15      &    17$^{\rm h}19^{\rm m}15^{\rm s}$  &    63$^{\circ}10^{\prime}04{\arcsec}$  &   $30  \pm        3  $                    & 46.5   &   3.3   &   9.9  \\
   16      &    17$^{\rm h}22^{\rm m}28^{\rm s}$  &    63$^{\circ}32^{\prime}03{\arcsec}$  &   $47  \pm        10 $                    & 44.2   &   3.0   &   13.2  \\
   17      &    17$^{\rm h}29^{\rm m}21^{\rm s}$  &    63$^{\circ}54^{\prime}58{\arcsec}$  &   $17  \pm        8  $                    & 50.4   &   2.3   &   22.1  \\
   18      &    17$^{\rm h}26^{\rm m}13^{\rm s}$  &    64$^{\circ}07^{\prime}35{\arcsec}$  &   $39  \pm        6  $                    & 34.5   &   3.3   &   17.8\\
   19A, 19B &   17$^{\rm h}28^{\rm m}23^{\rm s}$  &    64$^{\circ}22^{\prime}18{\arcsec}$  &   $25  \pm        1, 20  \pm        4  $  & 96.5, 63.8   &   7.33, 3.4   &   20.8  \\
%  19B     &    17$^{\rm h}28^{\rm m}23^{\rm s}$  &    64$^{\circ}22^{\prime}18{\arcsec}$  &   $20  \pm        4  $                    &   \\
   20A, 20B&    17$^{\rm h}16^{\rm m}25^{\rm s}$  &    65$^{\circ}29^{\prime}27{\arcsec}$  &   $34  \pm        4, 36  \pm        5  $  & 21.1, 16.9   &   3.1, 1.9    &8.7  \\
%  20B     &    17$^{\rm h}16^{\rm m}25^{\rm s}$  &    65$^{\circ}29^{\prime}27{\arcsec}$  &   $36  \pm        5  $                    &    \\
   21      &    17$^{\rm h}20^{\rm m}38^{\rm s}$  &    65$^{\circ}19^{\prime}09{\arcsec}$  &   $37  \pm        12 $                    & 12.0   &   1.3   &   12.1  \\
   22      &    17$^{\rm h}20^{\rm m}04^{\rm s}$  &    65$^{\circ}02^{\prime}16{\arcsec}$  &   $67  \pm        16 $                    & 4.9    &   1.2   &   10.7  \\
   23A, 23B&    17$^{\rm h}14^{\rm m}13^{\rm s}$  &    65$^{\circ}09^{\prime}43{\arcsec}$  &   $34  \pm        7, 47  \pm        9  $  & 55.0, 24.3   &   3.0, 1.7   &   5.8   \\
%  23B     &    17$^{\rm h}14^{\rm m}13^{\rm s}$  &    65$^{\circ}09^{\prime}43{\arcsec}$  &   $47  \pm        9  $                    &    \\
   24      &    17$^{\rm h}15^{\rm m}30^{\rm s}$  &    64$^{\circ}39^{\prime}47{\arcsec}$  &   $38  \pm        6  $                    & 14.8   &   1.8   &   4.4  \\
   25      &    17$^{\rm h}10^{\rm m}33^{\rm s}$  &    64$^{\circ}30^{\prime}02{\arcsec}$  &   $18  \pm        7  $                    & 19.5   &   1.9   &   3.9  \\
   26      &    17$^{\rm h}08^{\rm m}02^{\rm s}$  &    64$^{\circ}04^{\prime}24{\arcsec}$  &   $35  \pm        9  $                    & 64.0   &   1.2   &   6.7  \\
   27A,27B &    17$^{\rm h}06^{\rm m}10^{\rm s}$  &    64$^{\circ}38^{\prime}11{\arcsec}$  &   $20  \pm        9, 10  \pm        12 $  & 29.2, 22.0   &   1.8, 1.4   &   9.6   \\
%  27B     &    17$^{\rm h}06^{\rm m}10^{\rm s}$  &    64$^{\circ}38^{\prime}11{\arcsec}$  &   $10  \pm        12 $                    &    \\
   28      &    17$^{\rm h}05^{\rm m}03^{\rm s}$  &    64$^{\circ}24^{\prime}08{\arcsec}$  &   $18  \pm        8  $                    & 5.6   &   1.2   &   10.8  \\
   39      &    17$^{\rm h}02^{\rm m}11^{\rm s}$  &    65$^{\circ}01^{\prime}03{\arcsec}$  &   $18  \pm        6  $                    & 7.5   &   2.6   &   15.4  \\
   30      &    17$^{\rm h}13^{\rm m}17^{\rm s}$  &    63$^{\circ}48^{\prime}19{\arcsec}$  &   $63  \pm        19 $                    & 133.2 &   8.0   &   1.8  \\
\noalign{\smallskip}
\hline    
\noalign{\smallskip}
\multicolumn{7}{l}{$^a$ Total intensity peak flux}\\
\multicolumn{7}{l}{$^b$ Polarized intensity peak flux}\\
\multicolumn{7}{l}{$^c$ Distance from cluster center}
\end{tabular}                          
}                                      
\end{center}                           
\end{table*}

\subsection {The 30 background sources}
\label{the30sources}

\begin{figure}
\begin{center}
\includegraphics[scale=.85]{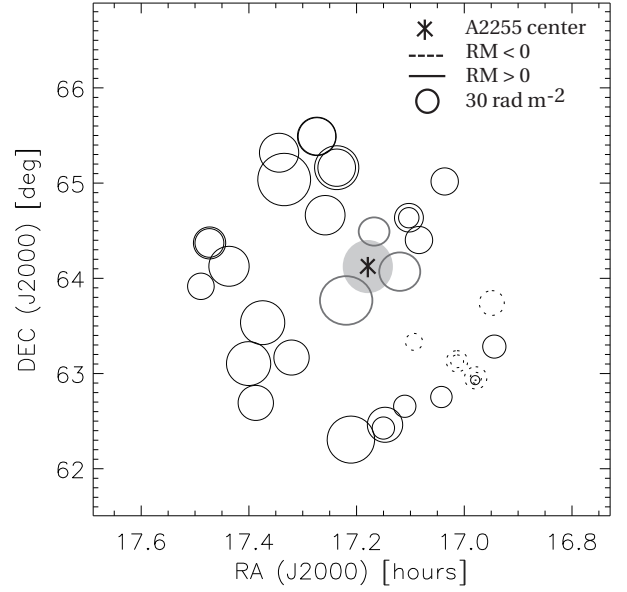}
\caption{The RM of the 30 sources observed within 2$^{\circ}$ from the
center of \object{A2255}. Each circle is centered on the location of a source
and its area is proportional to the RM value. In the legend, a scale
factor is also shown. The gray circle indicates the
X-ray boundaries of \object{A2255} as derived from the {\it ROSAT} PSPC observation of the cluster in the 0.5-2.0~keV band \citep{2003AJ....125.2427M}.}
\label{30sourcesfig}
\end{center}
\end{figure}

\begin{figure}
\begin{center}
\includegraphics[scale=.85]{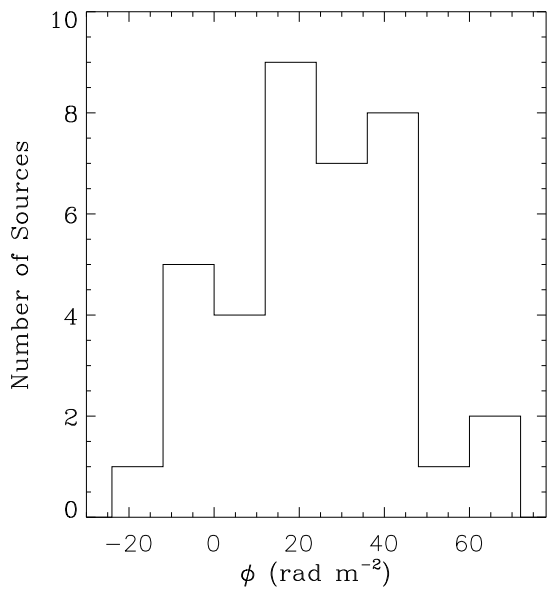}
\caption{Histogram of the RM distribution of the 30 background sources
binned in intervals of 12~rad~m$^{-2}$.}
\label{rmhist}
\end{center}
\end{figure}

Table \ref{30sourcestable} lists the results of the analysis
of the 30 background sources observed around \object{A2255}. Seven out of the
30 sources have a double structure\footnote {The NVSS has a
45${\arcsec}$ resolution while the WSRT 21~cm PSF is
$\sim12{\arcsec}\times13{\arcsec}$.}. In this case, we computed the RM
of each component (labeled A and B in Table
\ref{30sourcestable}). The histogram of the RM values for the
30 sources is presented in Fig.~\ref{rmhist} and Fig.~\ref{30sourcesfig} shows the results. At the location
of each source, we plot a circle, whose area is proportional to the RM
value of the source. Under the assumption that the RM of the sources
is representative of our Galaxy, we can conclude that in the field of
\object{A2255} the Galactic foreground fluctuates on large scales showing
values between 0~rad~m$^{-2}$ to +40~rad~m$^{-2}$, going from SW to
NE. This agrees with the RM depths at which the Galactic
foreground is detected in the RM cube at 85~cm (see
Sect.~\ref{RMcubeat85cm}). Moreover, the RM values of each
component of the double sources are in close agreement. This suggests
that along these lines of sight our Galaxy does not have fluctuations
on scales smaller than 15${\arcsec}$.\\  
By fitting a Gaussian profile to
the RM distribution of the sources located within 1 degree of the cluster center, we conclude that the Galactic
foreground at the location of \object{A2255} ($\rm l=94^{\circ}, \rm b= 35^{\circ}$) has a mean Faraday depth of
approximately $+37 \pm 8$~rad~m$^{-2}$ with a dispersion $\sigma_{\rm RM} = 19$~rad~m$^{-2}$. 

%30 sources, we obtain a $<\rm RM> = 24 \pm 4$~rad~m$^{-2}$ and a dispersion $\sigma_{\rm RM} = 22$~rad~m$^{-2}$. 
% Since there is evidence of a gradient in the Galactic foreground in the field of the cluster in our 85 cm RM cube (see Sect.~\ref{RMcubeat85cm}), we decide to limit limit our analysis to the sources located within a radius of
%$1^{\circ}$ from the center of \object{A2255}. By doing that, we conclude that the Galactic
%foreground at the location of the cluster ($\rm l=94^{\circ}, \rm b= 35^{\circ}$) has a mean Faraday depth of
%approximately $+37 \pm 8$~rad~m$^{-2}$ with a dispersion $\sigma_{\rm RM} = 19$~rad~m$^{-2}$.} 

\subsection{The RM cubes}
\label{TheRMcubes}

The identification of real astronomical structures and instrumental
artifacts in the RM cubes is done efficiently by animated scanning
through them.  The RM cubes, as well as the original complex
polarization images, show many structures and patterns that have an
instrumental origin and are not associated with the real cluster
synchrotron emission. These are a combination of uv plane and
image-plane effects. Multiplicative errors in the uv plane result in a
convolution with an error pattern in the image plane. Therefore, the
strongest effects are associated with the strongest sources, although
they may extend over a substantial part of the image plane. In the
following subsections we present the real features detected in the
final RM cubes, discuss their nature, and investigate the origin
of the instrumental artifacts. It is important to remember that
intensity in an RM cube is expressed in mJy~beam$^{-1}$~RMSF$^{-1}$
\citep{br2005}.

\subsubsection{RM cube at 18+21+25~cm}
\label{RMcubeat18+21+25cm}

A total of 543 complex polarization images were used in constructing
the RM cube\footnote{\object{A2255}\_18\_21\_25CM.gif, only available in electronic form at the CDS via anonymous ftp to cdsarc.u-strasbg.fr (130.79.128.5) or via http://cdsweb.u-strasbg.fr/cgi-bin/qcat?J/A+A/} in this wavelength range. Each
input image covers a field of view of 1.5$^{\circ} \times$
1.5$^{\circ}$. A range of Faraday depths from --1000 to
+1000~rad~m$^{-2}$ was synthesized, with a step of
10~rad~m$^{-2}$. The input images to RM-synthesis cover a considerable
range of frequencies, ranging from 1.15~GHz to 1.78~GHz. We brought
them to the same resolution of the 25~cm data by applying a Gaussian
taper to the data. The RM cube has an rms-noise level of
8.5~$\mu$Jy~beam$^{-1}$. To increase the signal-to-noise ratio for the
extended low brightness structures, we also made an RM cube at half
resolution ($28{\arcsec} \times 30{\arcsec}$). Its noise is
approximately 10$~\mu$Jy~beam$^{-1}$.

\begin{figure}
\begin{center}
\includegraphics[scale=.43]{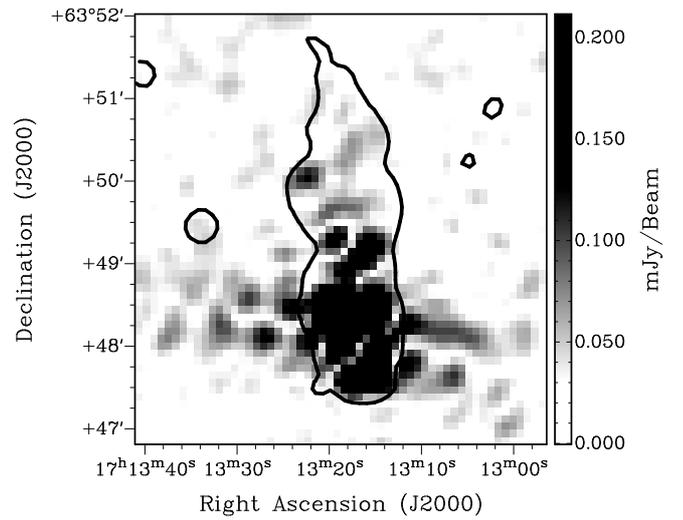}
\caption{Error pattern at the location of the Beaver radio galaxy in
the high-frequency RM cube at $\phi$~=~+30~rad~m$^{-2}$ (see
text). The polarized intensity is in units of
mJy~beam$^{-1}$~RMSF$^{-1}$. The contour is that of the total
intensity 25~cm map and is at 0.1~mJy~beam$^{-1}$ (${\rm FWHM} =
14{\arcsec} \times 15{\arcsec}$).}
\label{beaverspikes}
\end{center}
\end{figure}

The main instrumental artifacts in this high-frequency RM cube are
related to the presence of spikes associated with the strongest
sources in the field. The effect is most visible at the location of
the Beaver radio galaxy (RA~=~17$^{\rm h}13^{\rm m} 19^{\rm s}$,
DEC~=~$+63\hbox{$^{\circ}$}48\hbox{$^{\prime}$
}16\hbox{$\arcsec$}$). The intensity of these spikes rapidly decreases
when we move away from the east-west direction. Their peak intensity
is about 0.1~mJy~beam$^{-1}$ (see
Fig.~\ref{beaverspikes}). Their cause is still unclear.

\begin{figure*}
\begin{center}
\includegraphics[scale=.6]{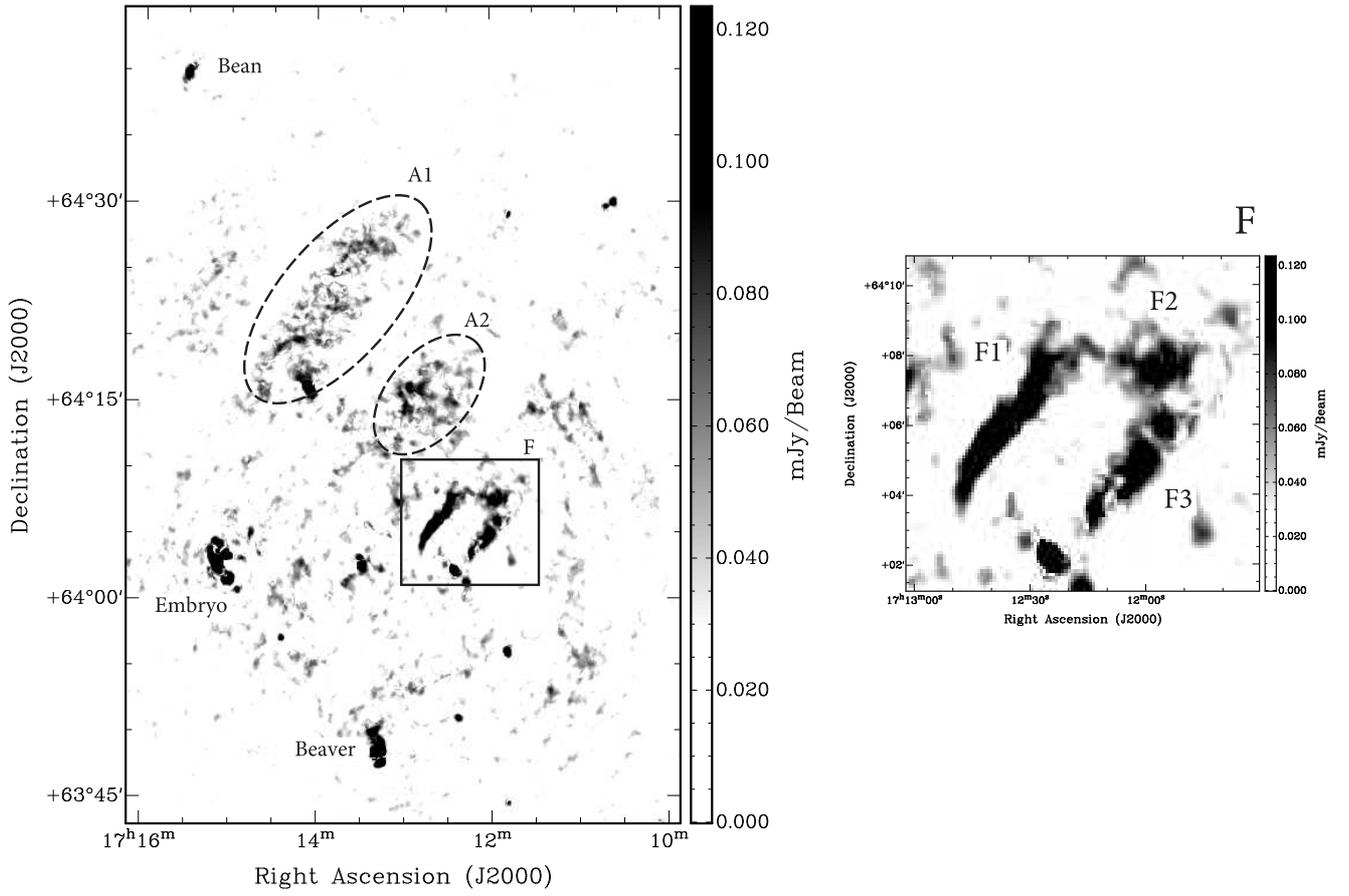}
\caption{Top panel: polarized intensity (in units of
mJy~beam$^{-1}$~RMSF$^{-1}$) in the field of \object{A2255} from the high-frequency RM cube (18~cm + 21~cm + 25~cm) at
$\phi$~=~+30~rad~m$^{-2}$. Bottom panel: zoom into the region where
the radio filaments are located.}
\label{RM252118_frame103}
\end{center}
\end{figure*}

As an example of the real astronomical signal detected in the field
of \object{A2255} in this RM cube, in Fig.~\ref{RM252118_frame103} we show the
frame at $\phi$~=~+30~rad~m$^{-2}$. We can distinguish two main
components of polarized signal:
\begin{itemize}
\item the signal associated with cluster structures, i.e. radio
  galaxies and filaments. This component is imaged better in the full
  resolution RM cube and extends between $-200$~rad~m$^{-2}$ and +200
~rad~m$^{-2}$, with each structure having its own Faraday depth;
\item two large-scale polarized structures (labeled A1 and A2 in
Fig.~\ref{RM252118_frame103}). That these components have no
counterpart in total intensity favor an association with the Galactic
foreground \citep{wier,2001ApJ...549..959G,haver2003,2009A&A...500..965B}. They appear at Faraday depths between $-20$~rad~m$^{-2}$ and +80~rad~m$^{-2}$ and peak at $\phi \sim 30$~rad~m$^{-2}$.
\end{itemize}

\begin{figure}
\begin{center}
\includegraphics[scale=.4]{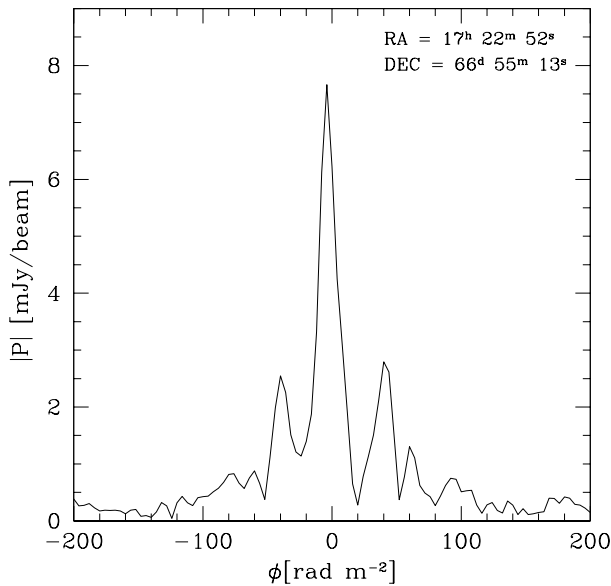}
\caption{Faraday spectrum of the off-axis source 4C +66.19 extracted
from the 85~cm RM cube. The flux in in units of
mJy~beam$^{-1}$~RMSF$^{-1}$. The resonances at the value $\mid \phi
\mid$ $\sim42$~rad~m$^{-2}$ come from to the 17~MHz periodicity of the Q
and U signals for off-axis sources (see text). RM-cleaning has been
applied.}
\label{42rad}
\end{center}
\end{figure}

\subsubsection{RM cube at 85~cm}
\label{RMcubeat85cm}
\label{foregroundpage}
 
The complex polarization images at 85~cm only included baselines
 longer than 300~m. This made it possible to remove most of the large-scale Galactic polarized emission. The RM cube\footnote{\object{A2255}\_85CM.gif, only available in electronic form at the CDS via anonymous ftp to cdsarc.u-strasbg.fr (130.79.128.5) or via http://cdsweb.u-strasbg.fr/cgi-bin/qcat?J/A+A/} was produced by
 combining 400 complex polarization images, each of them covering a
 field of view of $6^{\circ} \times 6^{\circ}$. The RM cube
 synthesizes a range of Faraday depths from $-400$~rad~m$^{-2}$ to
 +400~rad~m$^{-2}$, with a step of 4~rad~m$^{-2}$. The angular
 resolution within the frequency range 310--380~MHz varies by a factor
 of 1.22. To obtain an almost frequency-independent beam, a Gaussian
 taper was applied to the data. The RM cube has an rms noise of
 45~$\mu$Jy ~beam$^{-1}$, the lowest achieved at such a long
 wavelength to date. However, this level is only reached at the edge
 of the field and/or at high RM values ($\mid$$\phi$$\mid$ $>$
 100~rad~m$^{-2}$), where instrumental artifacts are less severe.

The main instrumental artifacts in the 85~cm RM cube are related to
 residual off-axis polarization. Radio emission
observed at large distance from the pointing center of the telescope
is generally instrumentally polarized. The component that is uniform
across the field of view is small (a few \%) and is calibrated away
through standard procedures (see
Sect.~\ref{bandpassandpolarisationcalibration}). The
instrumental polarization in parabolic dishes, however, rapidly
increases with distance from the optical axis
\citep[e.g. ][]{1999ASPC..180...37N}. In the case of the WSRT, the
off-axis polarization at 85~cm causes spurious signals at the location
of many strong sources in the field. The narrow spread in off-axis
polarization levels for the different telescopes in the array also
causes ring-like patterns.\\ At $\phi$~=~0 rad
m$^{-2}$, all contributions add up coherently, thus if the off-axis
polarization were frequency-independent, the response in the
RM cube should rapidly decrease at high RM values. However, the WSRT
has a strong frequency dependence with a period of 17~MHz, which
causes peaks in the Faraday spectrum at values of about~$\pm$42
rad~m$^{-2}$. Figure~\ref{42rad} shows this for a source
$\sim2^{\circ}$ northeast of the cluster center.

Despite the spurious signal, the RM cube shows bright polarized
emission ~with an~astronomical ~origin. This ~~is ~~mostly ~~detected ~~in
~the ~RM range --20~rad~m$^{-2}$ $\leq$ $\phi$ $\leq$ +50~rad~m$^{-2}$
and is likely from our Galaxy, since it extends well beyond the
cluster ``boundary'', and it has no counterpart in total intensity. It
shows two morphologically distinct components.
\begin{itemize}
\item The first one, detected in the range
--20~rad~m$^{-2}$~$\leq$~$\phi$ $\leq$~+20~rad~m$^{-2}$, is
elongated along the north-south direction, and has a polarization
angle that varies mostly on large scales.
\item The second one is detected in the range +20 rad~m$^{-2}$ $\leq$
$\phi$ $\leq$ +50~rad~m$^{-2}$ and it is oriented along the NE to SW
direction, has a complex structure, and shows very rapid changes in the
polarization angle.
\end{itemize}
The physical properties of the Galactic foreground as detected in the
field of \object{A2255} will be described and analyzed in detail in
\citet{pizzogalactic}.

\begin{figure*}[!]
\begin{center}
\includegraphics[scale=.64]{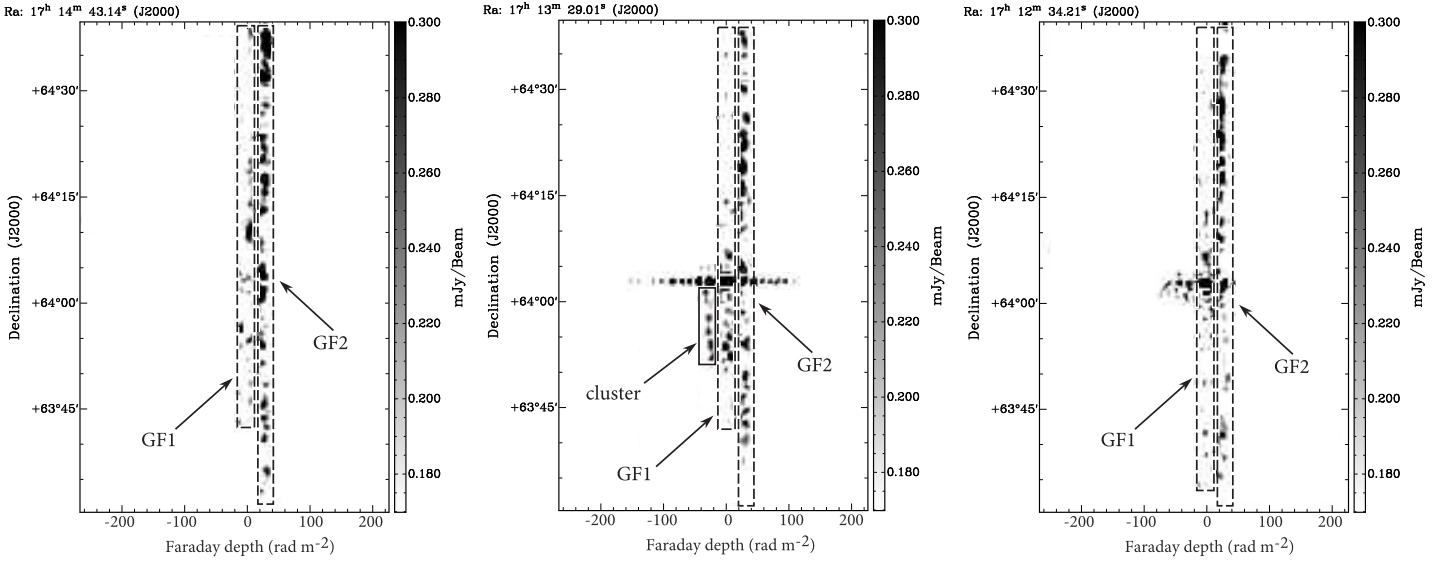}
\vspace{0.05cm}
\caption{The polarized emission in the 85~cm RM cube in units of
  mJy~beam$^{-1}$~RMSF$^{-1}$. From left to right, the
  situation at decreasing values of right ascension is presented. The
  right ascension value is reported in the upper left corner of each
  panel. At RA~=~17$^{\rm h}13^{\rm m} 29^{\rm s}$ three polarized
  components are detected, nominally the cluster emission
  (``cluster'') and the two Galactic Faraday screens (GF1 and
  GF2). These two components are the only ones detected at the other
  right ascensions. The emission at location
  DEC~=~+64\hbox{$^{\circ}$}02\hbox{$^{\prime}$ }43\hbox{$\arcsec$}
  and extending in Faraday depth between --70~rad~m$^{-2}$ to
  +100~rad~m$^{-2}$ comes from the Double radio galaxy.}
\label{decprofilezoom}
\end{center}
\end{figure*}

\begin{figure*}[!]
\begin{center}
\includegraphics[scale=.65]{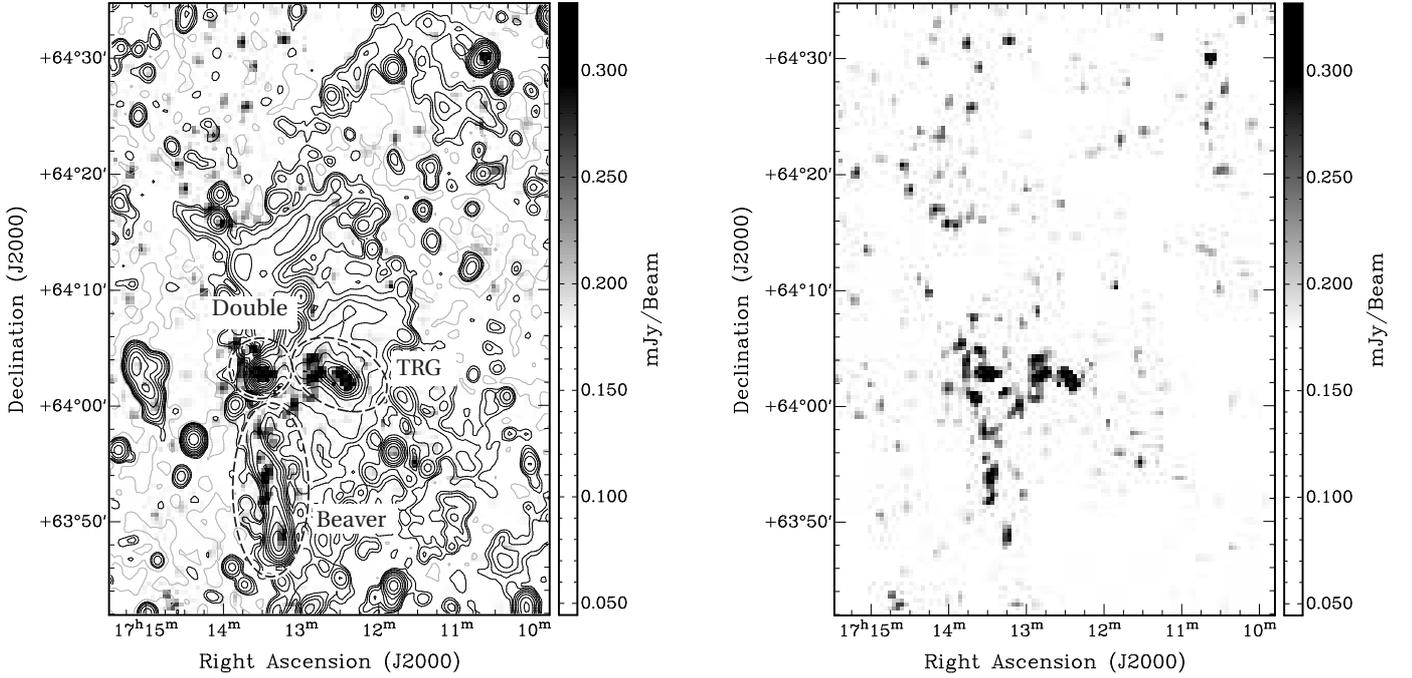}
\vspace{0.05cm}
\caption{The polarized emission (in units of
mJy~beam$^{-1}$~RMSF$^{-1}$) in the field of \object{A2255} at 85~cm is shown
with (left panel) and without (right panel) total intensity contours
on top. The frames refer to a Faraday depth of $\phi$=
--24~rad~m$^{-2}$ and have a noise level of about 30~$\mu$Jy~beam$^{-1}$. The noise level of the total intensity image is $\sim$0.1~mJy~beam$^{-1}$. The contours of the toal intensity image are --0.3 (gray), 0.3, 0.6, 1.2, 2.4,
4.8, 9.6, 20, 40, 80, 160~mJy~beam$^{-1}$. The resolution is
54${\arcsec} \times 64{\arcsec}$.}
\label{85cmpolarizationcomb}
\end{center}
\end{figure*}

Although most of the polarized flux in the RM cube comes from our Galaxy, there
are features that suggest an association with continuum structures belonging
to the cluster. This emission focuses in the southern regions of the cluster
and between declinations +63\hbox{$^{\circ}$}51\hbox{$^{\prime}$
}48\hbox{$\arcsec$} and +63\hbox{$^{\circ}$}59\hbox{$^{\prime}$
}33\hbox{$\arcsec$}. To show this, we produced images of the polarized
emission at the location of the cluster as a function of the right ascension
(see Fig.~\ref{decprofilezoom}). The central panel of Fig.~\ref{decprofilezoom} refers to RA~=~17 $^{\rm h}13^{\rm m} 29^{\rm
  s}$. Here we can distinguish 3 components of polarized emission in the
field. Two of them are located at $\phi$ $\sim0$~rad~m$^{-2}$ and at positive
Faraday depths (GF1 and GF2) and represent the two screens due to
our Galaxy. The last one is visible at Faraday depths ranging between
--32~rad~m$^{-2}$ and --16~rad~m$^{-2}$ and is only detected at this specific
value of right ascension. This is confirmed by the left and the right panels
of Fig.~\ref{decprofilezoom}, where the situation at right ascensions
to the east and to the west of the cluster is presented.\\ Inspecting the
RM cube at the Faraday depths at which the cluster component is detected, we
note that this is associated with the tail of the Beaver radio galaxy, which
is undetected in total intensity and polarization at higher frequencies. In
Fig.~\ref{85cmpolarizationcomb} we plot the polarized emission at
$\phi = -24$~rad~m$^{-2}$ as an example. Given its characteristics, this
signal is real and not due to instrumental artifacts, which mainly show up at
$\phi$~=~0~rad~m$^{-2}$ and are mostly associated with point
sources. Additional significant polarized flux is visible in this frame at the
location of the Double and the TRG radio galaxies. This emission extends over
a wide range of Faraday depths, which agrees with the high-frequency
polarimetric results for these radio galaxies (see
Sect.~\ref{rotatiomeasurestructure}). Polarized emission is also
detected at the location of the Embryo radio galaxy at $\phi \sim+
32$~rad~m$^{-2}$, the same Faraday depth at which this radio galaxy is
detected in the high-frequency RM cube (see
Sect.~\ref{rotatiomeasurestructure}).

\subsubsection{RM cube at 2 m}
\label{RMcubeat2m}

The RM cube\footnote{\object{A2255}\_2M.gif, only available in electronic form at the CDS via anonymous ftp to cdsarc.u-strasbg.fr (130.79.128.5) or via http://cdsweb.u-strasbg.fr/cgi-bin/qcat?J/A+A/} at 2~m has
been produced by adding 1523 complex polarization images, which cover
a field of view of 3.5$^{\circ} \times 3.5^{\circ}$.  The RM cube
synthesizes a range of Faraday depths from --100~rad~m$^{-2}$ to
+100~rad~m$^{-2}$, with a step of 0.5~rad~m$^{-2}$. Within the
frequency range 115--175~MHz, the beam sizes vary by a factor of 1.5,
so the data have been tapered to obtain an almost frequency-independent beam. The final rms noise level in the RM cube should be
$\sim0.7$~mJy~beam$^{-1}$. However, the presence of significant
residual secondary lobes associated with Cas~A and Cyg~A results in a
higher noise level in the final cube
($\sigma\sim1$~mJy~beam$^{-1}$). Although these sources were
subtracted during the imaging, their residual contribution is still
present, owing to the difficulty of modelling their flux density and
morphology. In the 2~m RM cube, there is no evidence of any polarized
signal associated with \object{A2255} or with our Galaxy.

\section{The radio galaxies and the filaments}
\label{theradiogalaxiesandthefilaments}

In the following subsections, we review the properties of the six
radio galaxies to which we limited our analysis and of the filaments, while reporting
their polarization percentages. The Sidekick radio galaxy was left out of the
analysis because of its small linear size. The polarization properties of the
radio sources at 18,
21, and 25~cm are summarized in Table
\ref{polarisationpercentages}. The data at 85~cm and 2~m were
excluded because of strong Galactic foreground emission (85~cm) or
not detecting of the sources of interest in polarization (2~m).

The polarized surface brightness of the sources was estimated by integration
over all the Faraday depths in which they are detected in the RM cube. The
result was then corrected for the RM cube noise level. The analytical
expression of this procedure is given by
\begin{equation}
\mid\mid P \mid\mid = B^{-1} \sum_{i=1}^n \left( \mid\mid F(\phi_i)\mid\mid -\sigma \sqrt{\frac{\pi}{2}} \right) \;,
\end{equation}
where $B$ is the~ area~ under~ ~the ~restoring beam ~of the ~RM-CLEAN
~divided by~ $\Delta\phi=\mid\phi_{i+1}-\phi_i\mid$, and $\sigma
\sqrt{\frac{\pi}{2}}$ the average value of the ricean noise
distribution of $\mid\mid F(\phi)\mid\mid$ \citep{Brentjens}.\\ Since
we do not correct for the off-axis instrumental polarization, we
expect that the measured polarization percentages are more affected by
instrumental polarization for those sources farther away from the field
center.

\begin{table}
\caption{Fractional polarization ($m$) and depolarization ($DP$) for the radio galaxies and the
filaments.}
\label{polarisationpercentages}
\smallskip
\begin{center}
{\small
\begin{tabular}{ccccccc}
\hline
\hline
\noalign{\smallskip}
Source     &   $m_{18 cm}$ $^a$  &   $m_{21 cm}$   &   $m_{25 cm}$   &   DP$_{18 cm} ^{21 cm}$ $^b$   &   DP$_{21 cm} ^{25 cm}$   &      DP$_{18 cm} ^{25 cm}$  \\
           &       \%       &       \%        &       \%        &                           &                           &                             \\
\noalign{\smallskip}
\hline
\noalign{\smallskip}
Double     &   2.4           &      1.8       &       1.6        &          0.8             &           0.9             &              0.7            \\
\noalign{\smallskip}
\hline
\noalign{\smallskip}
Goldfish   &   3.7           &      2.2       &       1.6        &          0.6             &           0.7             &              0.4            \\
\noalign{\smallskip}
\hline
\noalign{\smallskip}
TRG        &   2.1           &       2.0      &       1.4        &          1             &           0.7             &              0.7            \\
\noalign{\smallskip}
\hline
\noalign{\smallskip}
Bean       &   -             &       18       &       17         &          -               &            0.9            &               -             \\
\noalign{\smallskip}
\hline
\noalign{\smallskip}
Embryo     &   17            &       19.5     &        19       &          1.1             &            1            &                1.1            \\
\noalign{\smallskip}
\hline
\noalign{\smallskip}
Beaver     &   16.1          &        13.3     &        11.4     &           0.8            &            0.9           &                0.7           \\
\noalign{\smallskip}
\hline
\noalign{\smallskip}
F1         &    29.9         &        27.6     &       21.0      &           0.9            &            0.8           &                0.7           \\
\noalign{\smallskip}
\hline
\noalign{\smallskip}
F2         &    41.3         &        37.8     &       37.0      &           0.9            &             1            &                 0.9          \\
\noalign{\smallskip}
\hline
\noalign{\smallskip}
F3         &    44.0         &        39.4     &       39.4      &           0.9              &             1            &                 0.9            \\          
\noalign{\smallskip}
\hline
\noalign{\smallskip}
\noalign{\smallskip}
\multicolumn{7}{l}{$^a$ Defined as the ratio between the polarized and the total}\\
\multicolumn{7}{l}{~~~intensity flux.}\\
\multicolumn{7}{l}{$^b$ The depolarization between $\lambda_1$ and $\lambda_2$, where $\lambda_1 > \lambda_2$,}\\ 
\multicolumn{7}{l}{~~~ is defined as $\frac{m_{\lambda_1}}{m_{\lambda_2}}$.}
\end{tabular}
}
\end{center}
\end{table}

\begin{figure*}[!]
\begin{center}
\includegraphics[scale=0.77]{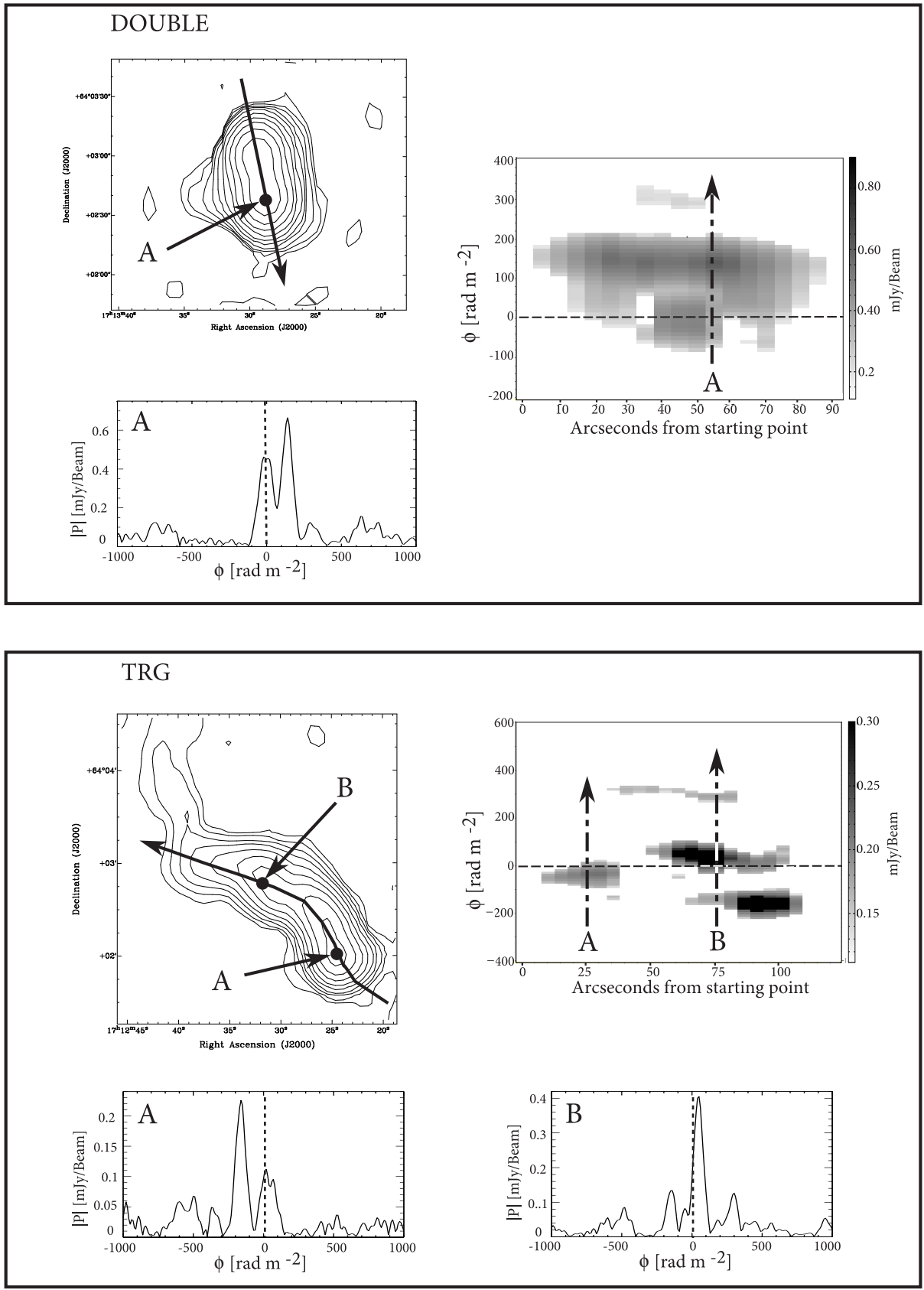}
\caption{Polarimetric properties of the six radio galaxies belonging
to \object{A2255} derived from the high-frequency RM cube. The panels are
labeled with the source name and show the 25 cm total intensity
contour map plus one or two Faraday spectra extracted at the specified
locations (A and B). The arrow passing through each radio galaxy
represents the direction along which we extracted the profile of the
polarized emission through the RM cube, presented in the right area of
each panel. Here, the polarized intensity is in units of
mJy~beam$^{-1}$~RMSF$^{-1}$. The dotted line in the Faraday spectra is
at $\phi$~=~0~rad~m$^{-2}$. The resolution of the total intensity map
is 14${\arcsec} \times$ 15${\arcsec}$ and the contours are at 0.1,
0.2, 0.4, 0.8, 1.6, 3.2, 6.4, 12, 24, 48~mJy~beam$^{-1}$.}
\label{spectraradiogalaxies_I}
\end{center}
\end{figure*}

\addtocounter{figure}{-1}
\begin{figure*}[!t]
\begin{center}
\includegraphics[scale=0.77]{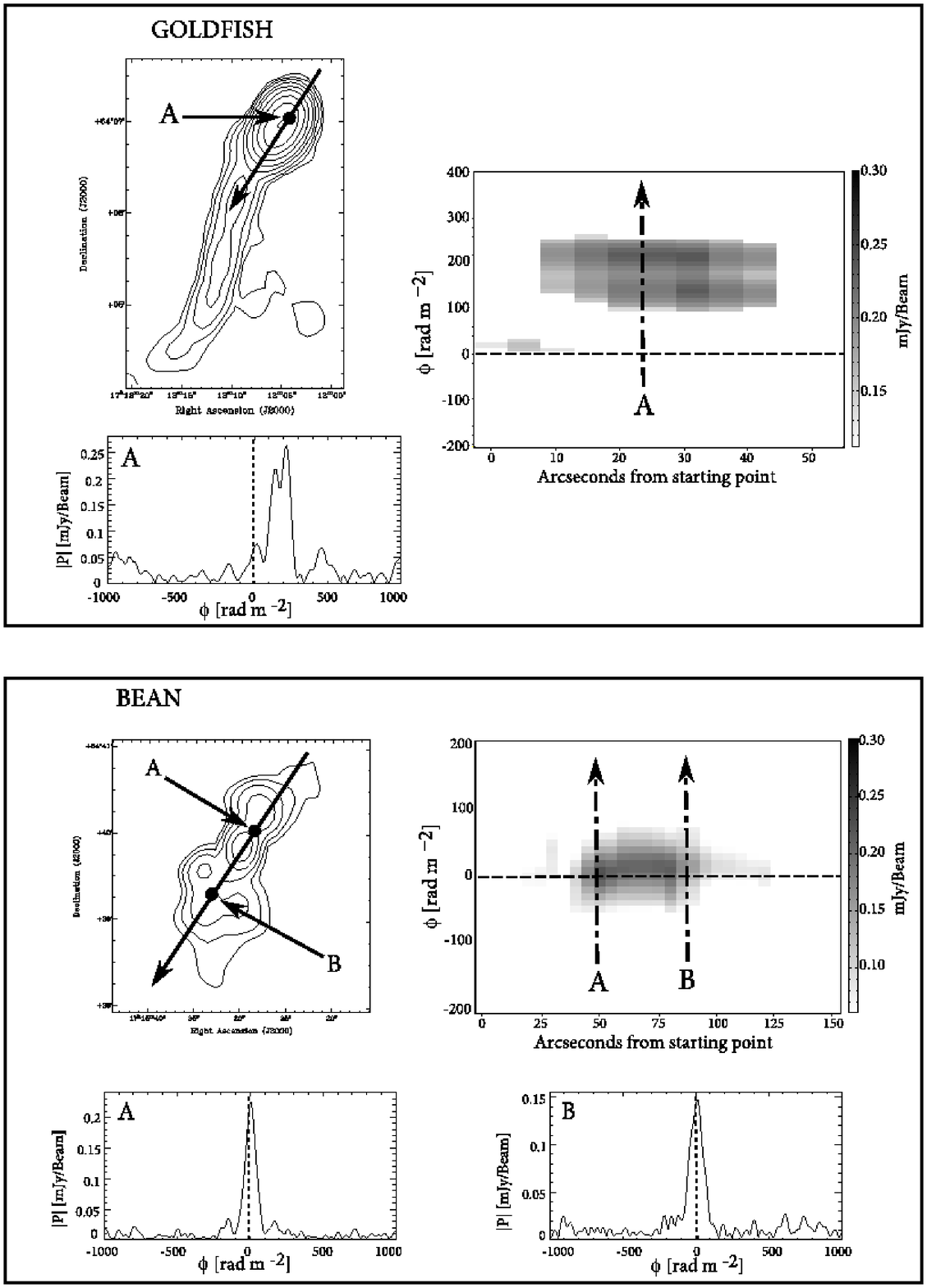}
\caption{Continued.}
\label{spectraradiogalaxies_II}
\end{center}
\end{figure*}

\addtocounter{figure}{-1}
\begin{figure*}
\begin{center}
\includegraphics[scale=0.77]{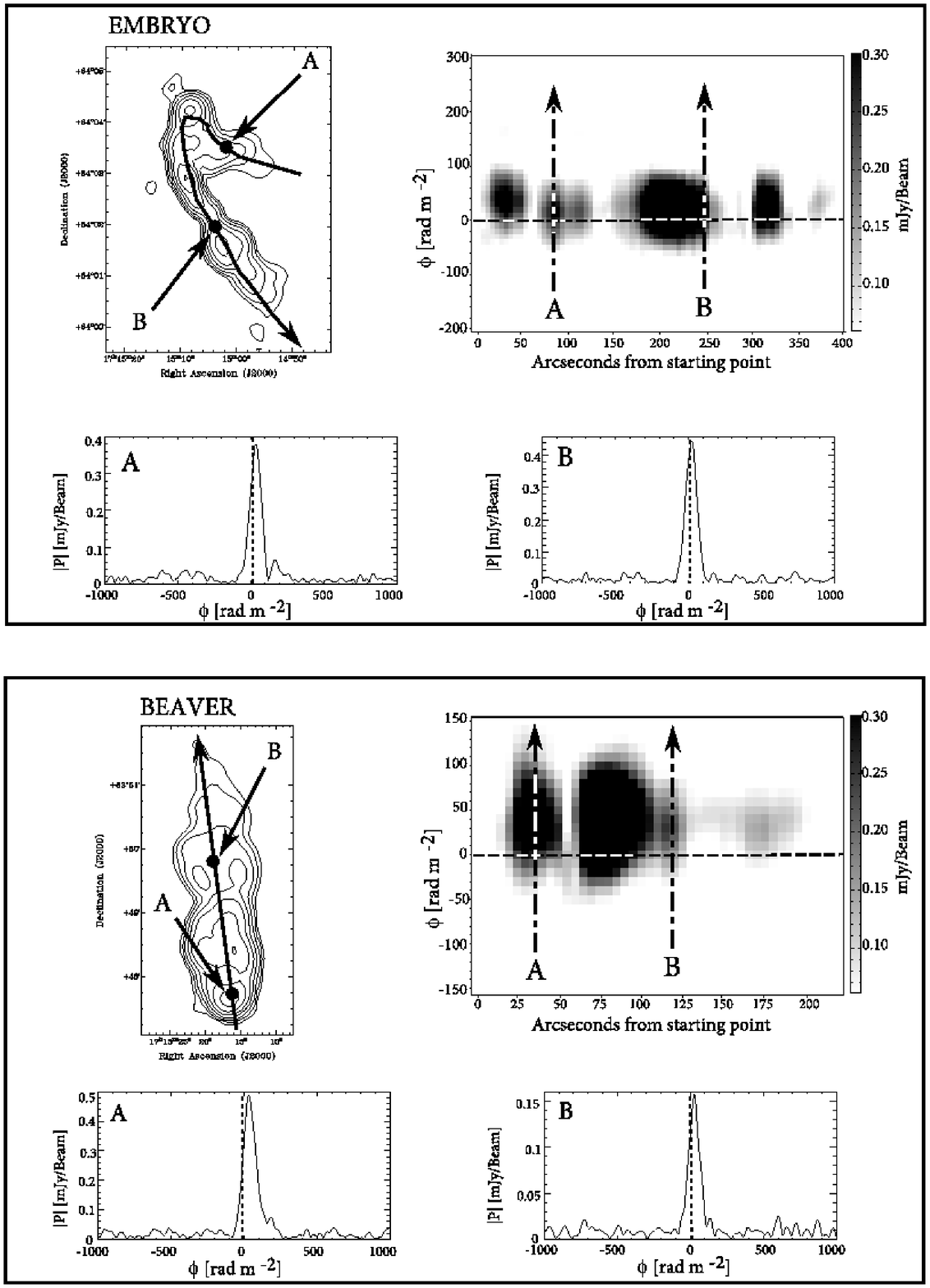}
\caption{Continued.}
\label{spectraradiogalaxies_III}
\end{center}
\end{figure*}

\begin{figure}[!]
\begin{center}
\includegraphics[scale=0.83]{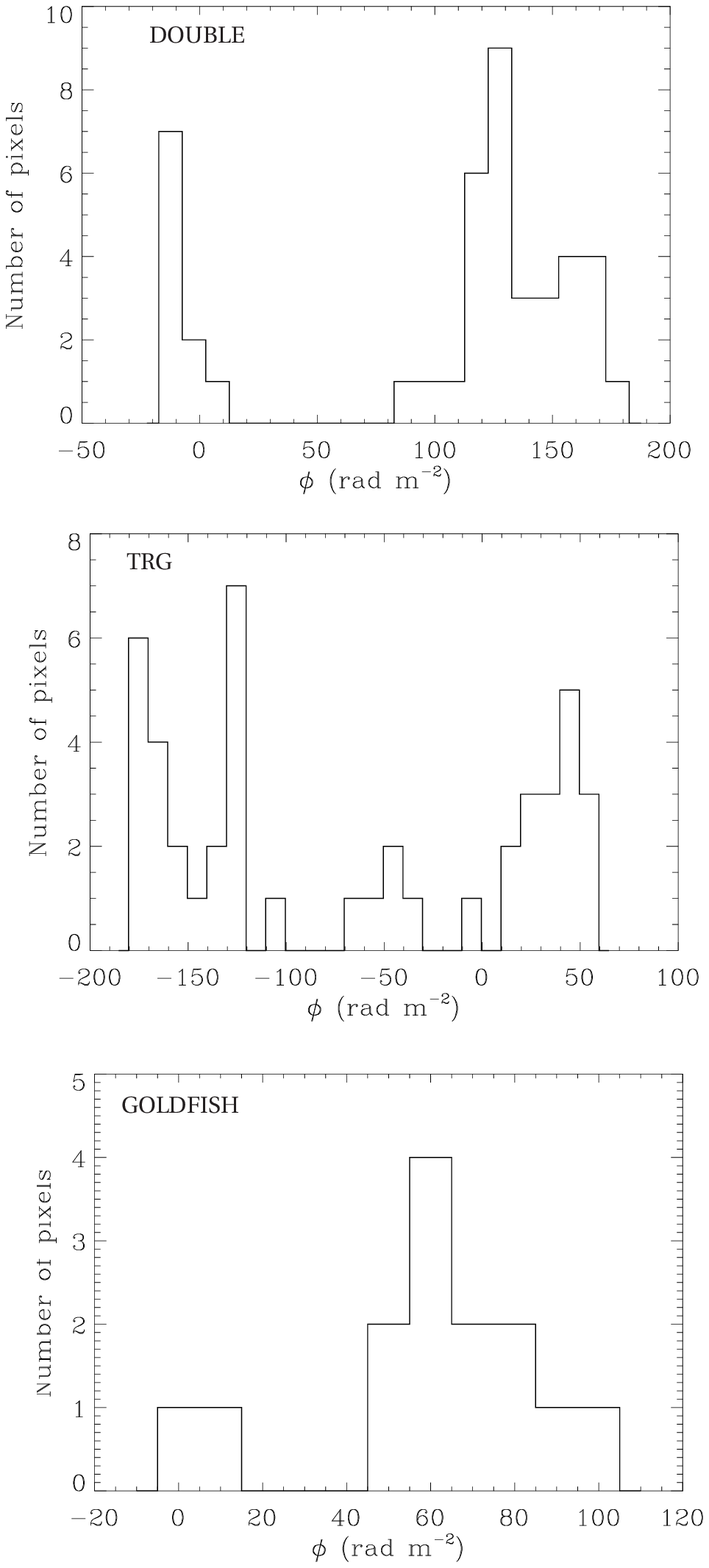}
\caption{RM distributions of the central radio galaxies of \object{A2255}.}
\label{histogramscentralradiogalaxies}
\end{center}
\end{figure}

\subsection{The Double}
\label{thedouble}

This radio galaxy has a double structure and a total extent of
63${\arcsec}$ (95~kpc). \citet{fer} noted that the Double lies exactly
in the merger region of \object{A2255}, suggesting that the lack of distortion in the
morphology could be related to being a young source
triggered by the merger. This radio galaxy is polarized at the 2.4\% level
at 18~cm, at 1.8\% level at 21~cm, and at 1.6\% level at 25~cm.

\subsection{The Goldfish}
\label{thegoldfish}
This NAT radio galaxy is located in projection near the cluster
center. From the nucleus two jets extend towards the southwest,
giving rise to a tail with an angular extent of $\sim174{\arcsec}$
(261~kpc). The tail bends to the south at a distance of 42${\arcsec}$
(63~kpc) from the head and then towards southeast at a distance of
120${\arcsec}$ (180~kpc). In the high-frequency RM cube, only the
nucleus of the source is polarized at a level of 3.7\%, 2.2\%, and 1.6\% at
18, 21, and 25~cm, respectively.

\subsection{The original TRG}
\label{theoriginaltrg}

This radio galaxy has an NAT morphology, with a tail extent of $\sim206{\arcsec}$ (300~kpc). Starting from the head, the tail is directed
towards the cluster center, and it bends towards north after 109${\arcsec}$ (160~kpc). The source is polarized at 2.1\%, 2.0\%, and 1.4\% at
18, 21, and 25~cm, respectively.

\subsection {The Bean}
\label{thebean}

This radio galaxy lies at 3.5~Mpc from the center of \object{A2255}. Given its
peripheral location, no detailed studies of this source exist in
the literature. The Bean shows a tailed morphology, but it is not
clear whether it should be classified as an NAT or a WAT \citep[wide
angle tail, ][]{1976ApJ...203L.107R}, given its apparent inclination
with respect to the line of sight. Its maximum angular extent is
$\sim78{\arcsec}$ (110~kpc), and it is polarized at 17\% at 21~cm and 17\% at 25~cm. These values are significantly higher than
those of the other extended radio galaxies of \object{A2255}. Point sources
(expected to be instrumentally polarized) lying at approximately the
same projected distance from the cluster center show a much lower
fractional polarization. Therefore, we conclude that the computed
fractional polarization of the Bean should be mostly intrinsic and not
instrumental. Since this source lies outside the field of view at
18~cm, it is not possible to give an estimate of its fractional
polarization at this wavelength.

\subsection{The Embryo}
\label{theembryo}

The Embryo lies at $\sim1.6$~Mpc from the cluster center and has a WAT
morphology. Its angular extent is $\sim4^{\prime}$ (360~kpc) at
25~cm. Twin jets originate from the core and extend to the northeast
and southwest. The former bends towards the cluster center at a
distance of $\sim40{\arcsec}$ from the head, while the latter remains
straight. This source is polarized at a 17\%, 19.5\%, and 19\% at 18,
21, and 25~cm, respectively.

\subsection{The Beaver}
\label{thebeaver}

The Beaver radio galaxy lies at $\sim1.6$~Mpc from the cluster center
and has a NAT morphology. Its size dramatically changes between high
and low frequency. A recent study of this radio galaxy in total
intensity \citep{pizzospectrum} shows that the tail increases its
length to almost 1~Mpc between 25~cm and 85~cm. This suggests very
steep spectral index values for the ending part of the tail
($\alpha_{25 \rm cm} ^{85 \rm cm} < -2.2\pm0.2$ and $\alpha_{85 \rm
cm} ^{2 \rm m} < -3.3\pm0.2$ , $S(\nu) \propto \nu^{\alpha}$),
where the relativistic electrons have suffered large energy losses
after their ejection from the parent galaxy.  At 25~cm the total
extent of the Beaver is $\sim240{\arcsec}$ (360~kpc), and it is
polarized at 16.1\%, 13.3\%, and 11.4\% at 18, 21, and 25~cm,
respectively.

\subsection{The filaments}
\label{thefilaments_1}

In our high-frequency observations, the three filaments are clearly detected
both in total intensity and polarization. In the following, we refer to
them using the convention adopted in Fig.~1 in \citet{gov}. They lie near the
cluster center and are located at the edges of the halo, which 
therefore has an uncommon rectangular shape. Between 18~cm and 25~cm, their
morphology and size do not change. The angular extent of F1, F2, and F3 is
323${\arcsec}$ (485~kpc), 330${\arcsec}$ (500~kpc), and 370${\arcsec}$ (550
kpc), respectively. Their fractional polarization ranges between 20\%--40\%
(see Table \ref{polarisationpercentages}) with an uncertainty of $\sim 2$\%.

\begin{figure*}[!htp]
\begin{center}
\includegraphics[scale=0.77]{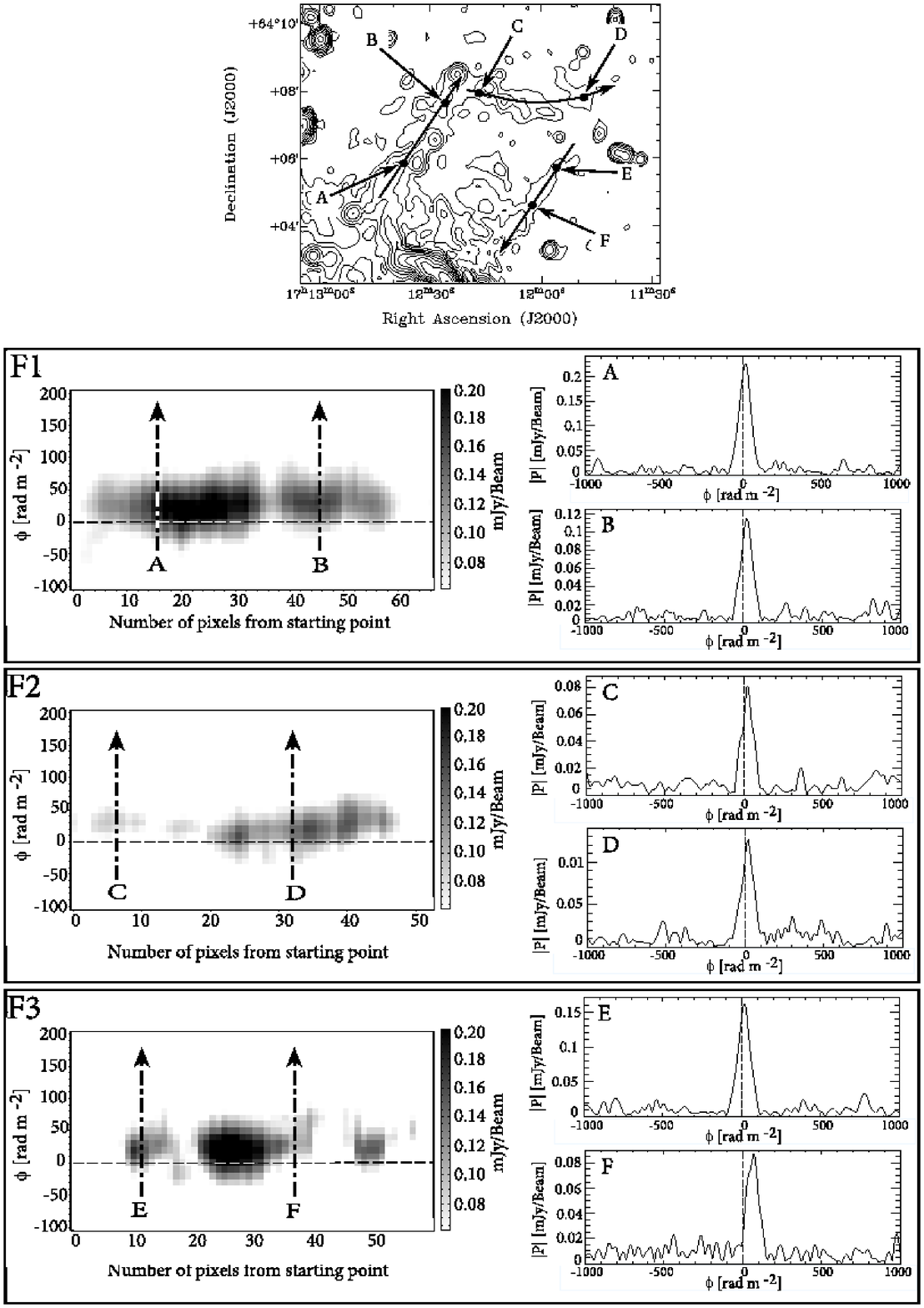}
\caption{The top panel presents the 25~cm total intensity map of the
filaments of \object{A2255}. The arrow passing through each filament represents
the direction along which the profile of the polarized emission though
the RM cube was extracted. These three profiles are reported in panels
F1, F2, and F3, together with a couple of examples of
Faraday spectra extracted at the locations specified in the total
intensity map (A, B, C, D, E, and F). The dotted line is at
$\phi$~=~0~rad~m$^{-2}$. The total intensity map has a resolution of
14${\arcsec} \times$ 15${\arcsec}$ and the contours are at 0.05, 0.1,
0.2, 0.4, 0.8, 1.6, 3.2, 6.4, 12, 24~mJy~beam$^{-1}$.}
\label{spectrafilaments}
\end{center}
\end{figure*}

\subsection{Rotation measure structure}
\label{rotatiomeasurestructure}

To obtain the rotation measure maps of the radio galaxies and the filaments, we
produced masks of the sources from the total intensity image and applied
them to the high-frequency RM cube (18~cm + 21~cm + 25~cm). To increase the
signal-to-noise ratio for the weakest structures, for the analysis we decided
to use the RM cube at half resolution.

The RM value for each pixel within a source was obtained by fitting a
Gaussian profile to the observed RM distribution. For the brightest
sources in the field (Double, Goldfish, and TRG) the instrumental polarization is higher than the
thermal noise. For the Double radio galaxy, for
example, the instrumental polarization is $\sim80-90~\mu$Jy. Therefore, as
detection limit we have chosen 100~$\mu$Jy (10$\sigma$) for these sources, and 50~$\mu$Jy (5$\sigma$)
for the others.

The Faraday spectra of the radio galaxies and the filaments have
different levels of complexity. We show this property in
Fig.~\ref{spectraradiogalaxies_I}. In each panel we present the total
intensity image of one radio galaxy, including a few examples of
Faraday spectra extracted at the specified positions (A and B). The
profiles in Faraday space along one direction within the source
are also given. From this image it is evident that the sources that
lie in projection near the cluster center (Double, TRG, and Goldfish) have Faraday
spectra characterized by one main peak at a specific
Faraday depth, plus significant secondary peaks (above 5$\sigma$) at
different Faraday depths.  This property reflects on the complexity of
the RM distributions of these radio galaxies, that are characterized
by a complex and non-Gaussian profile (see
Fig.~\ref{histogramscentralradiogalaxies}). On the other hand, the
radio galaxies which lie at large projected distance from the cluster
center (Bean, Embryo, and Beaver in Fig.~\ref{spectraradiogalaxies_I})
and the radio filaments (see Fig.~\ref{spectrafilaments}) show Faraday
spectra with only one significant peak. For this reason the RM images
could only be produced for the three external radio galaxies and for
the 3 filaments.

\begin{figure*}[ph!]
\begin{center}
\includegraphics[scale=0.8]{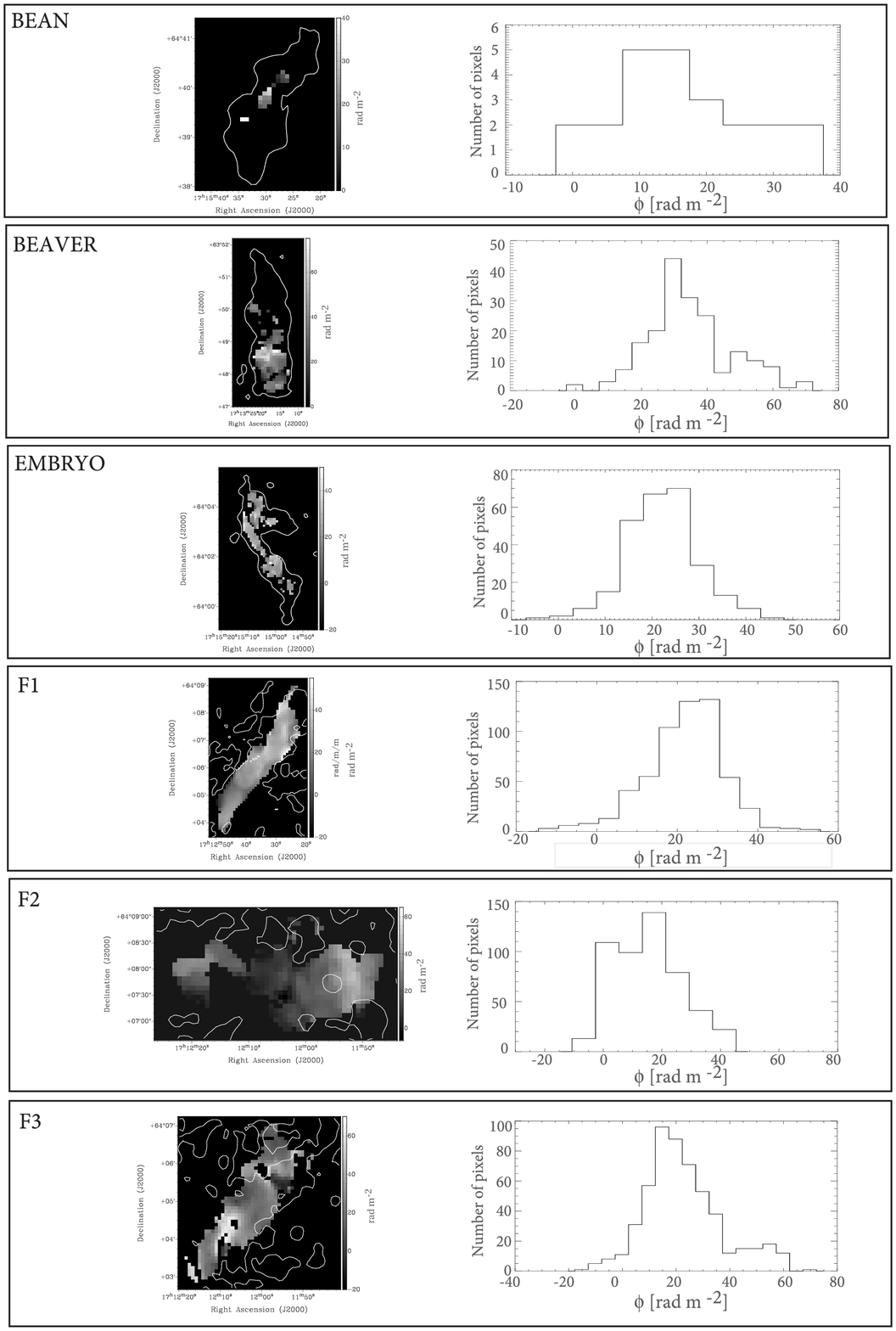}
\caption{Images of the rotation measure and histograms of the RM
distribution of the external radio galaxies and of the filaments,
obtained from the RM cube at high frequency. The images have a
resolution of 28${\arcsec} \times$ 30${\arcsec}$. The contours are drawn from the 25~cm map and are at 0.1~mJy~beam$^{-1}$.}
\label{rmimagescomposite}
\end{center}
\end{figure*}

Figure \ref{rmimagescomposite} presents the results. The RM
images have a resolution of $28{\arcsec} \times 30{\arcsec}$.  Besides
them, we plot the histograms of the RM distribution of the related
source. Table \ref{RMvalues} lists the mean $<$RM$>$ value,
$\sigma_{\rm RM}$, and the maximum absolute value of the RM
distributions. The radio galaxies and the filaments have similar
$<$RM$>$ and $\sigma_{\rm RM}$ values ($<$RM$> \sim +20$~rad~m$^{-2}$, $\sigma_{\rm RM}\sim10$~rad~m$^{-2}$). These values are typical
of radio sources lying at large projected distance from the cluster
center \citep[see Sect.~\ref{the30sources} and][]{2001ApJ...547L.111C} and are not significantly different from the expected Galactic foreground value in the direction of \object{A2255} (see Sect.~\ref{the30sources}).

\begin{table}
\caption{Parameters of the RM-distributions for the external radio
galaxies and for the filaments.}
\label{RMvalues}
\smallskip
\begin{center}
{\small
\begin{tabular}{ccccc}
\hline
\hline
\noalign{\smallskip}
Name       & $\mid$RM$_{max}$$\mid$    &    $<$ RM $>$       &   $\sigma_{\rm RM}$   &     Dist         \\
           &   rad~m$^{-2}$            &   rad~m$^{-2}$      &     rad~m$^{-2}$      &   r/r$_{c}$      \\
\noalign{\smallskip}
\hline
\noalign{\smallskip}
F1        &       52                   &          24         &           9           &      0.7          \\ 
\noalign{\smallskip}
\hline
\noalign{\smallskip}
F2        &       47                   &          14         &          13           &      1.2          \\ 
\noalign{\smallskip}
\hline
\noalign{\smallskip}
F3        &       70                   &          25         &          15           &      1.5          \\    
\noalign{\smallskip}
\hline
\noalign{\smallskip}
Embryo    &       47                   &          25         &           9           &      3.1          \\
\noalign{\smallskip}
\hline
\noalign{\smallskip}
Beaver    &       74                   &          37         &          13           &      4.1          \\
\noalign{\smallskip}
\hline
\noalign{\smallskip}
Bean      &       40                   &          19         &          11           &      7.5           \\
\noalign{\smallskip}
\hline
\end{tabular}
}
\end{center}
\end{table}

%\section {Discussion}
\label{discussion}

\subsection{The radio galaxies}
\label{theradiogalaxies}

The RM of extragalactic radio sources can be considered as the sum of the
contributions of three different regions, namely those internal to the source
itself, our Galaxy, and the ICM.
The contribution local to the source appears generally small, as shown by (i) the Laing-Garrington effect \citep{1988Natur.331..149L,1988Natur.331..147G,1991MNRAS.250..171G}, (ii) the radial trend observed in statistical Faraday studies \citep{2004JKAS...37..337C}, and (iii) the results of gradient alignment statistics \citep{2003ApJ...597..870E}. 
%
%
%The contribution local to the source
%appears generally small, based on the following arguments:
%\begin{itemize}
%\item observations of the Laing-Garrington effect, where one observes
%a different depolarization for the radio lobes of radio galaxies, with
%the lobe more distant from the observer being more depolarized
%\citep{1988Natur.331..149L,1988Natur.331..147G,1991MNRAS.250..171G};
%\item the radial trend observed in statistical Faraday studies
%\citep{2004JKAS...37..337C}, where point sources observed at large
%impact parameters from the cluster center have small RM values;
%\item the results of gradient alignment statistics
%\citep{2003ApJ...597..870E}, which suggest that there is no
%significant correlation between the intrinsic polarization angle of
%extended cluster radio galaxies and their RM maps.
%\end{itemize}
The contribution from our Galaxy is also small for objects lying at Galactic latitudes above b~$\sim20^{\circ}$  
\citep{1981ApJS...45...97S,2009ApJ...702.1230T}. Therefore, for
extragalactic objects at high Galactic latitude, the main contribution
to the observed Faraday rotation is represented by the intracluster
magnetic field.\\
The RM distribution of cluster radio
sources carries important information on the magneto-ionic
properties (electron density and magnetic field along the line of sight) of the external medium. If, in the simplest case, this is
characterized by a magnetic field that is tangled within cells of
uniform size and it has the same strength and random orientation
within them, the observed RM along any given line of sight is
represented by a random walk process. Therefore, the distribution of RM
results in a Gaussian with zero mean and the dispersion related to the
number of cells along the line of sight.

Several studies of the RM distribution of extended radio galaxies in
clusters have pointed out that there is a significant trend between
the observed RM distribution and the projected distance of the source
from the cluster center
\citep{1999A&A...344..472F,2001MNRAS.326....2T,2001A&A...379..807G}:
the smaller the projected distance from the core, the higher
$\sigma_{\rm RM}$ and $<$RM$>$. This result suggests that the external
Faraday screen for all the cluster sources is the ICM of the cluster,
which modifies the polarized radio signal depending on how much
magneto-ionized medium it crosses\footnote{It is worth noting that the RM
observed towards cluster radio galaxies may not be entirely
representative of the cluster magnetic field if the RM is locally
enhanced by the compression of the ICM from the motion of the source
through it. However, this hypothesis is ruled out because
\citet{2004JKAS...37..337C} showed that the RM distribution of point
sources seen at different impact parameters from the cluster center
has a broadening towards the center of the cluster. This result
reveals that most of the RM contribution comes from the ICM.}.

The polarimetric properties of  three radio galaxies belonging to \object{A2255} (the Double, the Original TRG, and the Beaver) have been studied by \citet{2006A&A...460..425G} by means of high-frequency (1.4~GHz) VLA observations of the cluster. The central sources (the original TRG and
the Double) show the highest $\sigma_{RM}$ and $<$RM$>$ of the sample,
while the peripheral radio galaxy (the Beaver) shows a low value of
$<$RM$>$ (= +36~rad~m$^{-2}$) and a comparable dispersion
($\sigma_{RM}$~=~42 rad~m$^{-2}$).  Our data confirm the
result for the Beaver and extend the analysis to other peripheral
radio galaxies (Embryo and Bean), which also show small $<$RM$>$ and
$\sigma_{RM}$. By combining our results with those of
\citet{2006A&A...460..425G}, it is evident that also in the case of
\object{A2255}, the external screen for the radio galaxies is the ICM. This affects not only the RM distribution of the cluster radio galaxies, but also the complexity of their 
Faraday spectra. Radio galaxies at
different projected distances from the cluster center show Faraday spectra with different levels of complexity. The sources
in the outermost cluster regions (Bean, Beaver, and Embryo) have
 simple spectra, mainly showing one peak, while the radio
galaxies located in the central areas of the cluster have complex
Faraday spectra, showing multiple
peaks. Since the co-location along the line of sight of multiple
Faraday regions produces Faraday spectra with multiple peaks
\citep{br2005}, we conclude that the radio galaxies showing the most
complex spectra (and accordingly the broadest RM distributions) are likely to be located deep inside the ICM or behind
the cluster. On the other hand, the radio galaxies showing the less complex spectra (and the smallest RM distributions) likely lie in the external regions of the ICM.\\ These results show that the RM distributions and the Faraday spectra of cluster radio galaxies are important for revealing their
3-dimensional location within the cluster.

\subsection {The Beaver radio galaxy}

\begin{figure}[h]
\begin{center}
\includegraphics[scale=0.55]{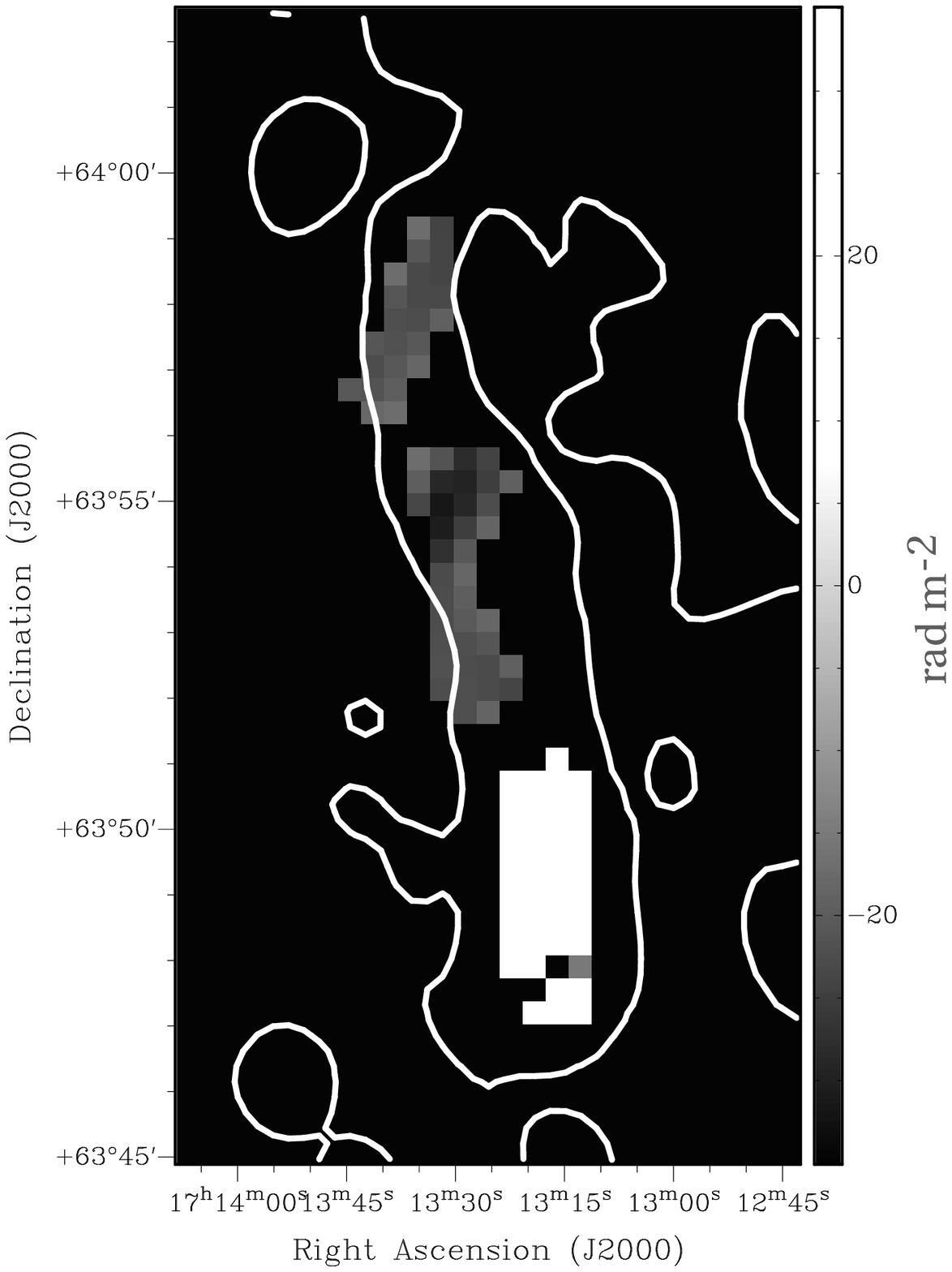}
\caption{RM image of the Beaver radio galaxy obtained by combining the
  information extracted from the high-frequency RM cube for the head and from
  the low-frequency RM cube for the tail. The contour is that of the 85~cm
  total intensity map (${\rm FWHM} = 54{\arcsec} \times 64{\arcsec}$) and is at
  1~mJy~beam$^{-1}$.}
\label{A2255RMBEAVERCOMPOSITE}
\end{center}
\end{figure}

\begin{figure}[h]
\begin{center}
\includegraphics[scale=0.3]{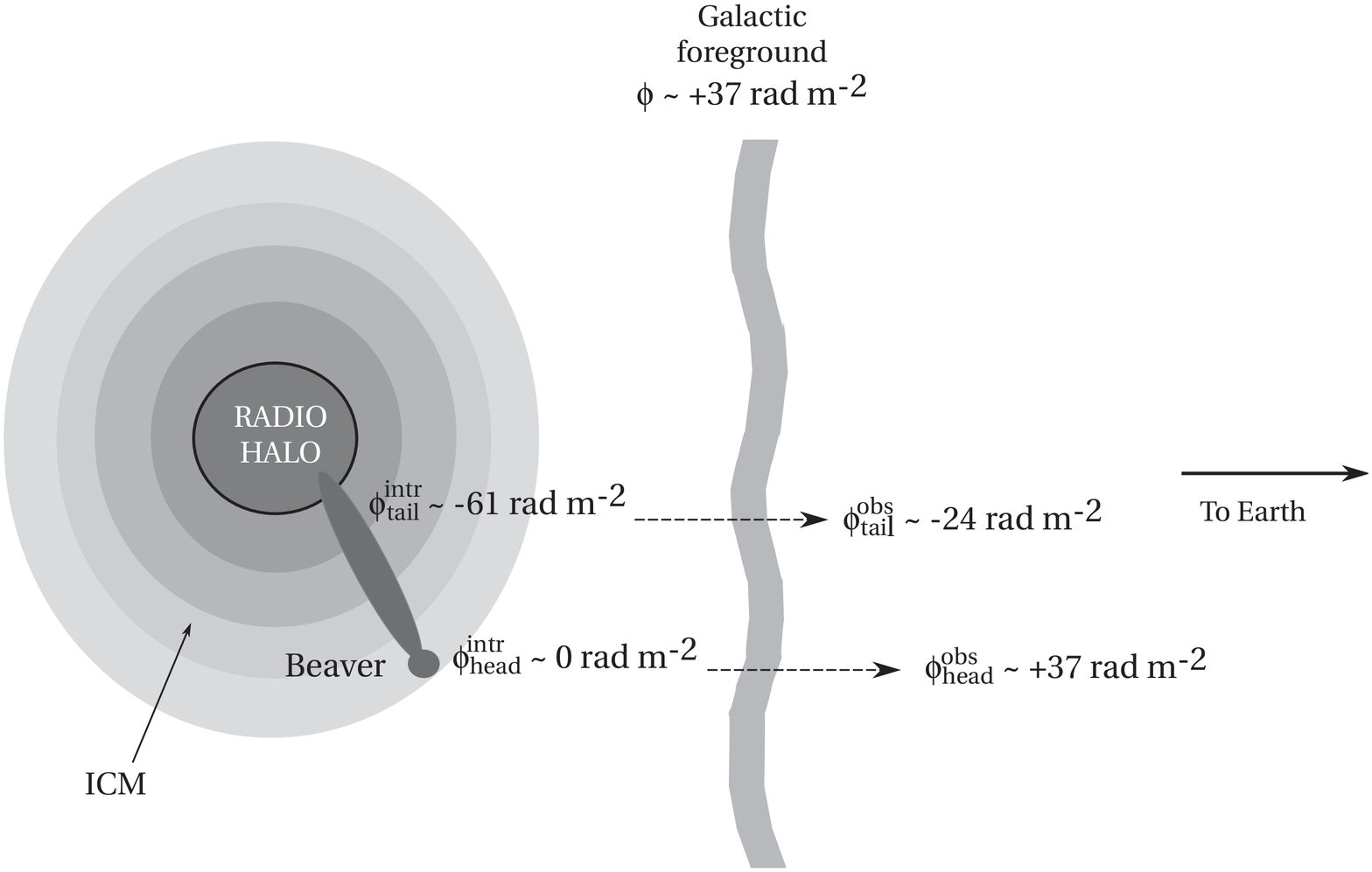}
\caption{Possible 3-dimensional location of the Beaver radio galaxy
  within \object{A2255}. The gray-scale for the ICM represents an increasing
  electron density of the plasma towards the cluster center. Our
  Galaxy contributes to the observed Faraday depths ($\phi_{head}
  ^{obs}$ and $\phi_{tail} ^{obs}$) at which the head and the tail
  appear in the RM cubes.}
\label{beaversketch}
\end{center}
\end{figure}

In the 85~cm RM cube, the tail of ~the ~Beaver ~radio ~galaxy ~is ~detected
between --32~rad~m$^{-2}< \phi <$--16~rad~m$^{-2}$. On the other hand, the
head and the initial part of the tail of the source are detected in the high-frequency RM cube at a Faraday depth of $\sim+37$~rad~m$^{-2}$, at which, in
the 85~cm RM cube, the emission mainly comes from the Galactic foreground. We
produced an RM map of the Beaver by combining the RM map of the head and the
initial part of the tail, as detected in the high-frequency RM cube (shown in
Fig.~\ref{rmimagescomposite}), and the RM map of the tail obtained
from the 85~cm RM cube. The latter has been produced by considering only the
frames corresponding to Faraday depths ranging between --32~rad~m$^{-2}$ and
--16~rad~m$^{-2}$ and using 5$\sigma$ (220~$\mu$Jy) as detection limit. The
image is presented in Fig.~\ref{A2255RMBEAVERCOMPOSITE}.\\ The clear
gradient between head and tail can give strong constraints on the possible
location of these two structures within \object{A2255}. It is worth noting that the
head and the initial part of the tail show RM values that are similar to
those of the Galactic foreground at the location of \object{A2255} (see
Sect.~\ref{the30sources}). This suggests that the head of the Beaver
is located on the outskirts of the cluster and, in particular, in the
foreground of \object{A2255}, where only a small portion of ICM is crossed by the radio
signal before reaching the observer. That the tail appears at
negative Faraday depths instead implies that it should be located deeper in
the ICM. Therefore, the Beaver could not lie in the plane of the sky, but with
the tail pointing towards the central radio halo, possibly connecting with
it. This interpretation is supported by the common spectral index values found
for the end of the tail and the southern region of the halo \citep{pizzospectrum}. The sketch in
Fig.~\ref{beaversketch} illustrates this situation.

\subsection{The filaments}
\label{thefilaments}

\label{filamentspage}
 
Our polarimetric data confirm the strong polarization of the filaments
and add important information that can help to explain their
nature. The filaments show similar polarimetric properties to those
found for the external radio galaxies of \object{A2255}, i.e. the Embryo, the
Beaver, and the Bean. Their RM distributions are characterized by low
values of $\sigma_{RM}$ and $<$RM$>$, with Faraday spectra showing
only one peak. Following the interpretation given for the radio
galaxies (see Sect.~\ref{theradiogalaxies}), these results
suggest that the filaments are not located deep in the ICM, but at the
periphery of the cluster. Moreover, given their high polarization
levels and their small spatial variance in RM (see
Fig.~\ref{rmimagescomposite}), we conclude that they should
be located in the foreground of the cluster and not in the
background. This is compatible with the polarization of the Galactic
foreground as detected in regions A1 and A2 in the high-frequency
RM cube (Fig.~\ref{RM252118_frame103}). The RM distribution
and the complexity of the spectra for regions A1 and A2 are similar to
those of the Bean (the most external radio galaxy belonging to \object{A2255})
and of the filaments. This indicates that most of the contribution to
the Faraday depth of these structures comes from our Galaxy and
suggests that the filaments should lie at large distance from the
cluster center. The observed central location of the filaments,
therefore, should be considered as due to a projection effect. {\it On
the basis of the elongated shape and the high degree of linear
polarization of the filaments, we therefore argue that they are relics
rather than part of a genuine radio halo.}

\subsection{The depolarization of the filaments}

At 85~cm and at 2~m we do not detect any polarized emission associated
with the radio filaments. At the Faraday depths at which the filaments
are detected in the high-frequency RM cube ($\phi\sim+30$~rad~m$^{-2}$), the polarized emission in the field of \object{A2255} in
the 85~cm RM cube is dominated by the Galactic foreground. This makes
it difficult to set a sensitive limit to their polarized emission. As
a conservative limit, we consider the polarized flux at the location
of these sources at the Faraday depth at which they are detected in
the high-frequency RM cube. This results in a fractional polarization
lower than 7.5\%. At 2~m, polarized emission associated with neighter the
cluster nor the Galactic foreground is detected in the
RM cube. Therefore, in this case, we can only set an upper limit to
the polarization of the filaments by considering 5 times the noise of
the RM cube itself. At the location of these structures, this
corresponds to a fractional polarization of 0.6\%.

To compare the polarized emission of the filaments at the different
frequencies, we smoothed the 21~cm data to the 85~cm angular
resolution. In the comparison we did not include the 2~m data because
their resolution is too low. By tapering the 21~cm data to
1$^{\prime}$ resolution, we see that the polarization of the filaments
decreases from more than 30\% to 14\%. The expected polarization of
the filaments at 85~cm should be well above the noise of the RM cube
at this wavelength. Therefore, we conclude that structure in the
magnetic field within these sources cannot be the cause of the
decrease of their fractional polarization. The observed depolarization
can be due to a variety of effects, discussed in the
following sections.

\subsection{Internal depolarization?}  

\begin{figure}[!]
\begin{center}
\includegraphics[scale=.4]{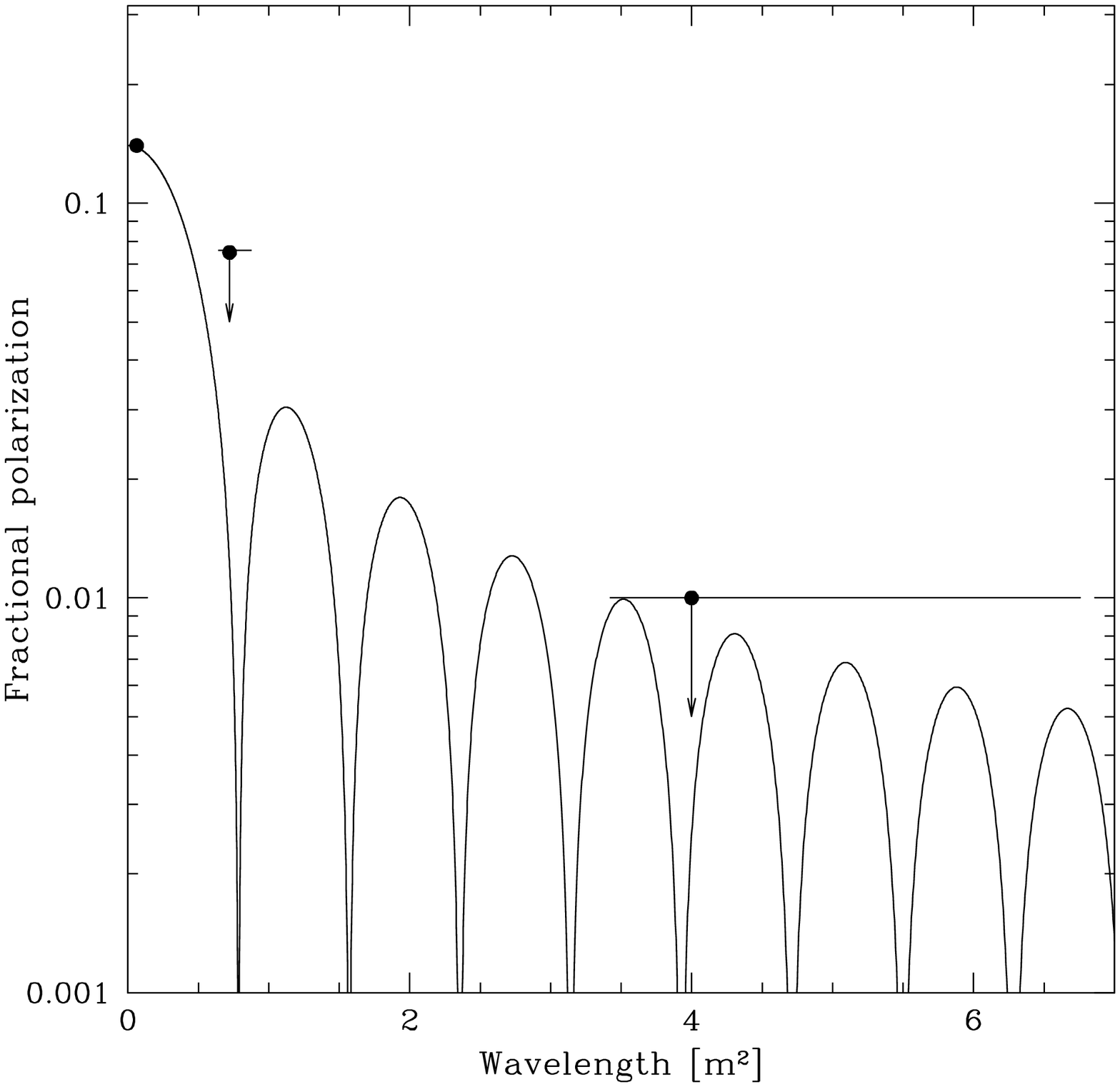}
\caption{Fractional polarization of the radio filaments. The points at 85 and
  200~cm are upper limits (see text). The 21~cm value was computed at the same
  resolution as the 85~cm observations. Given the low resolution of the
  2~m observations, the measurement at 150~MHz was not taken into account
  during the discussion. The horizontal bars indicate the $\lambda^2$ coverage
  of the observations. The sinc function corresponds to a Burn slab of Faraday
  thickness of 4~rad~m$^{-2}$, normalized to the 21~cm fractional polarization
  (14\%) at the 85 cm resolution.}
\label{slab}
\end{center}
\end{figure}

\citet{1966MNRAS.133...67B} studied in detail the case in which the
depolarization takes place within a radio source ({\it internal
depolarization}). In this situation, it is possible to determine the
Faraday thickness, or extent in Faraday depth, of the filaments. 
If the filaments have a uniform electron density and magnetic field,
their structure in Faraday space can be estimated by a slab model
\citep{1966MNRAS.133...67B}, which gives rise to the sinc function
displayed in Fig.~\ref{slab}. Our data agree with
a Faraday thickness of the filaments larger than 4~rad~m$^{-2}$.

\begin{figure}
\begin{center}
\includegraphics[scale=.4]{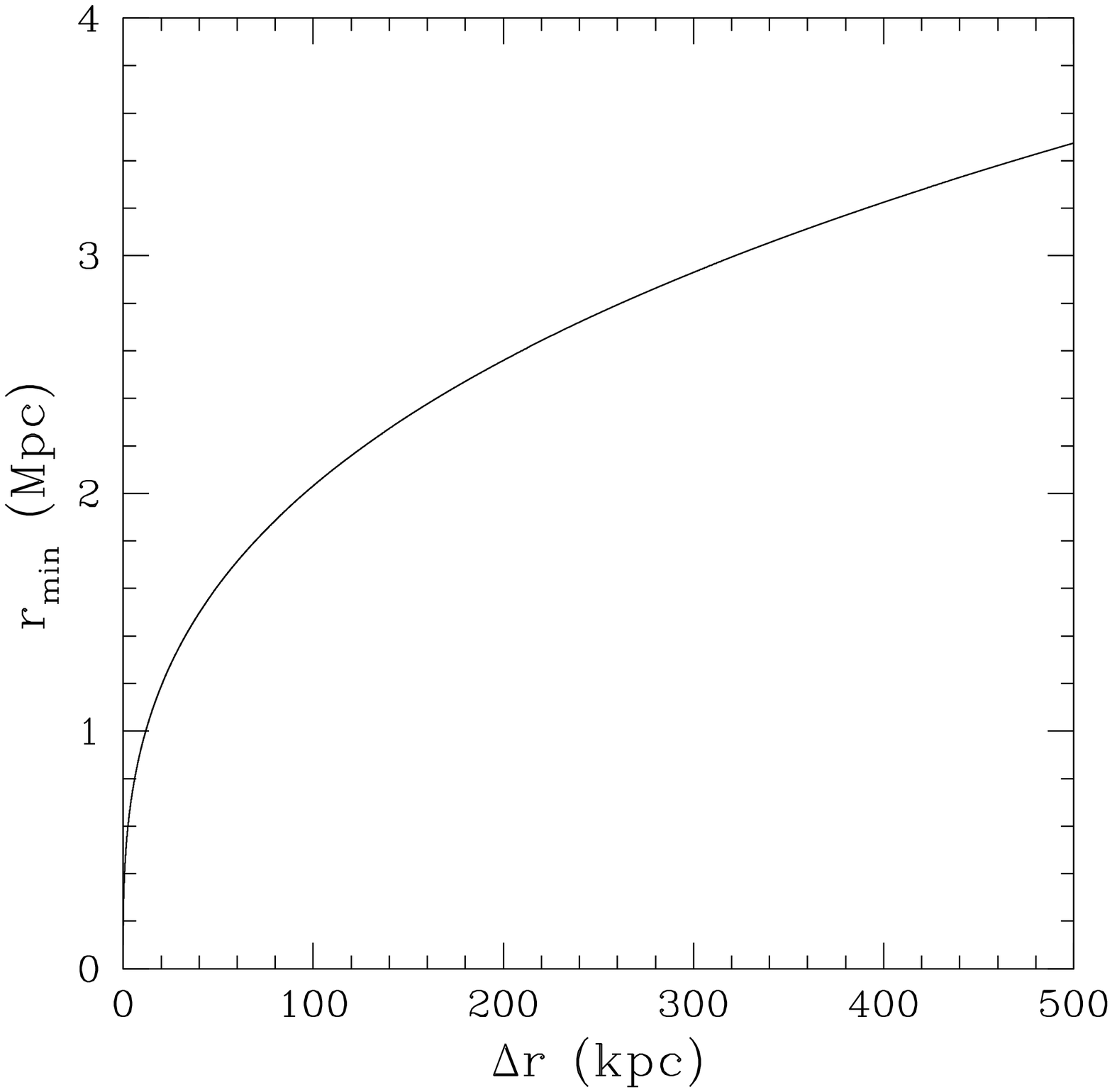}
\caption{Distance of the filaments from the cluster center ($r_{min}$)
as a function of their physical depth ($\Delta r$) on the assumption
that their Faraday thickness is larger than 4~rad~m$^{-2}$.}
\label{plot}
\end{center}
\end{figure}

With this result and with appropriate assumptions, it is possible to
derive an upper limit for the physical distance of the filaments from
the cluster center. The observed $\phi$ of these structures depends on
the electron density and on the magnetic field along the line of sight
through equation \ref{rmequation}. In the following
discussion we assume that the filaments have the same electron
density and magnetic field as the ICM at their location. However, it
is worth noting that most filaments are probably related to the
relics/ghosts studied by \citet{2001A&A...366...26E} and might have
much lower thermal densities than the external cluster plasma.\\ The
ICM electron density at a radial distance $r$ from the cluster X-ray
centroid is described by the standard $\beta$-model profile:
\begin{equation}
\label{electrondensity}
n(r)=\frac{n_0}{\left[1+ \left(\frac{r}{r_c}\right) ^2\right]^{3\beta /2}} \; ,
\end{equation}
where $r_c$ is the core radius and $n_0$ the central electron
density of the cluster. The gas density distribution of \object{A2255} was
derived by \citet{fer} and rescaled to our cosmology ($r_c$~=~432~kpc,
$n_0$~=~2.05 $\times$ 10$^{-3}$ cm$^{-3}$, $\beta$~=~0.74). \\
\citet{2006A&A...460..425G} found that the magnetic field of \object{A2255}
decreases from the cluster center to the periphery according to
\begin{equation}
\label{magnetic}
<{\bf B}>(r)=\frac{<{\bf B}>_0}{\left[1+ \left(\frac{r}{r_c}\right) ^2\right]^{3\mu /2}}\;,
\end{equation}  
where {\bf $<{\rm {\bf B}}>_0$} is the mean magnetic field at the
cluster center ($\sim$2.5~$\mu$G) and $\mu = \beta/2 = 0.37$ \citep{2006A&A...460..425G}.\\
Inserting Eqs.~\ref{electrondensity}-\ref{magnetic} into equation  \ref{rmequation}, we derive
\begin{eqnarray}
\label{rmform1}
\phi &<& 812 \int_{r_{min}} ^{r_{min}+\Delta r} \frac{n_0}{\left[1+ \left(\frac{r}{r_c}\right) ^2\right]^{3\beta /2}} \frac{<{\bf B}>_0}{\left[1+ \left(\frac{r}{r_c}\right) ^2\right]^{3\mu /2}} \it{d} r = \nonumber \\
&=& 812  \, n_0 \, B_0 \int_{r_{min}} ^{r_{min} +\Delta r} \frac{\it{d} r}
{\left[1+ \left(\frac{r}{r_c}\right) ^2\right]^{9 \beta /4}},
\end{eqnarray}
where $\phi$ is the adopted Faraday thickness of the filaments
(4~rad~m$^{-2}$), $r_{min}$ their minimum distance from the cluster
center, and $\Delta r$ is their physical depth. In the following
analysis, we assume $\beta=2/3\sim0.67$. Motivated by the
discussion in Sect.~\ref{thefilaments}, we assume
$R=r/r_c>> 1$; i.e, the filaments are located far away from the cluster
center. For example, at 3.5~Mpc distance ($r/r_c\sim8$), this
assumption gives an error below 10\%.\\ Equation
\ref{rmform1} then becomes

\begin{eqnarray}
\label{rmform3}
\phi         &<& 812  \, n_0 \, B_0 \int_{R_{min}} ^{R_{min} +\Delta R} \frac{\it{d} R}
{(1+ R^2)^{9 \beta /4}} = \nonumber \\
                 &=& 812 \, n_0 \, B_0 \,r_c\, \left[ - \frac {1}{2 R^2} \right]_{R_{min}} ^{R_{min} +
\Delta R} = \nonumber \\
                 &\simeq& 812 \, n_0 \, B_0 \,r_c \frac {\Delta R}{R_{min}^3}~.
\end{eqnarray}
In Fig.~\ref{plot} we plot the distance of the
filaments from the cluster center as a function of their physical
thickness. If we shape these structures as cylinders and assume
that their thickness (i.e. their width) is $\Delta R = 0.4$
\citep[$\sim180$~kpc, ][]{gov}, we find that they should lie within
approximately 6 core radii ($\sim2.6$~Mpc) from the cluster center to
be internally depolarized.

\subsection{External depolarization?}  
  
\begin{figure}
\begin{center}
\includegraphics[scale=.37]{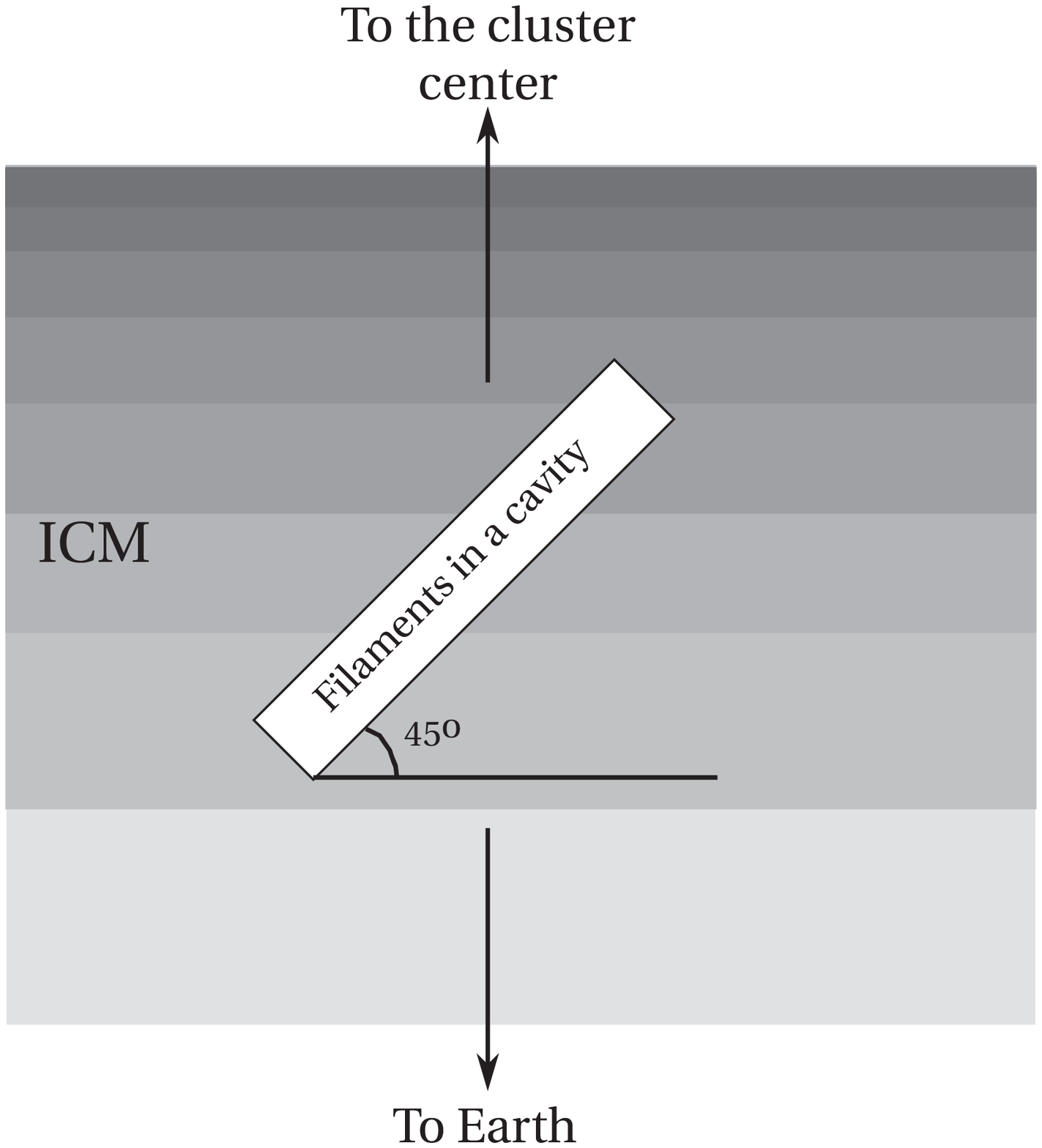}
\caption{View from the top of the filament region. The filaments
create a cavity in the ICM. Depolarization is produced only if
different regions of the filaments lie at different depths in the ICM
along the line of sight (see text). We assume that the filaments are
tilted by an angle of $45^{\circ}$ with respect to the plane of the
sky. The gray-scale represents the increasing electron density and
magnetic field strength towards the cluster center.}
\label{filaments45}
\end{center}
\end{figure}

\citet{1991MNRAS.250..726T} studied the case in which the
depolarization comes from a medium external to the radio source ({\it
external depolarization}).  This could be the ICM, in which the
filaments have expanded, creating a cavity.  Examples of such cavities
have been found in several clusters \citep[see
e.g. ][]{2004ApJ...607..800B,2006MNRAS.366..417F}. To date, they have
not been detected in \object{A2255}, but X-ray observations with a much higher
sensitivity than provided by the current X-ray satellites are
needed to possibly detect them. In the following, we assume the
ICM is uniformly stratified in terms of electron density and
magnetic field. Depolarization is only produced if different regions
of the filaments lie at different depths in the ICM along the line of
sight (see Fig.~\ref{filaments45}). The different remaining
path lengths through the cluster for the different parts of the
filaments will be sufficiently different to create a gradient in
Faraday depth. In fact, our 21 cm RM-images of the filaments reveal
 a gradient of $\sim5$~rad~m$^{-2}$~arcmin$^{-1}$ in
them. In this scenario, we can still apply equation
\ref{rmform3} to compute the upper limit for the distance of
these sources from the cluster center. The physical depth of the
filaments along the line of sight, which produces an RM gradient of
$\Delta R$=0.2 (90~kpc), corresponding to 1$^{\prime}$ resolution. In
this case, we find that the filaments should lie within 4.5 core
radii ($\sim1.9$~Mpc) from the cluster center to be depolarized.

Alternatively, the external depolarization could come from to an external
non-uniform Faraday rotating screen. At least two Faraday screens are likely to
be present between us and the filaments: one or more in the Galactic
foreground and one located at the cluster outskirts. In both scenarios, the observed depolarization of the filaments is
produced by the inhomogeneities of the foreground screen on scales
smaller than the observing beam. To depolarize a signal between 21~cm and 85~cm, fluctuations of
just 5~rad~m$^{-2}$~arcmin$^{-1}$ in a foreground screen are sufficient. However, such inhomogeneities are not observed in our Galaxy in the direction of \object{A2255}, where, instead, fluctuations on scales of  5~rad~m$^{-2}$~degree$^{-1}$ are more common \citep{2009ApJ...702.1230T,pizzogalactic}. This implies that not the Galactic foreground but a Faraday screen located at the cluster
outskirts could be responsible for the observed depolarization of the filaments. To
test this, low-frequency observations with higher resolution are needed.

\section{Summary and conclusions}
\label{summaryandconclusions}

We presented WSRT observations of the galaxy cluster Abell 2255 at 18,
21, 25, 85, and 200~cm. These were aimed at investigating the
polarimetric properties of the cluster radio galaxies and the uncommon
high polarization of the three radio filaments at the border of the
central halo. Using RM-synthesis, we produced RM cubes at the various
wavelengths. The high-frequency RM cube, obtained by combining the 18,
21, and 25~cm datasets, confirms that both the cluster radio galaxies
and the filaments are polarized. We have now also determined the RM of
the filaments. The RM cube at 85~cm is dominated by the polarized
emission associated with our Galaxy. However, there are several
features in it that argue for an association with continuum structures
belonging to the cluster.
The RM cube at 2~m, which has been produced after correcting the data
for the ionospheric Faraday rotation, does not show any polarized
emission associated with \object{A2255}.

Our polarimetric results at high frequency indicate that the cluster radio
galaxies located at a large projected distance from the cluster center
have the smallest $\sigma_{\rm RM}$ and $< \rm RM >$. The radial
decrease in the $< \rm RM >$, approaching the value of about
+30~rad~m$^{-2}$ (due to our Galaxy and also seen in the
background sources), is attributed to the ICM of \object{A2255}.
The radio galaxies lying at a small projected distance from the cluster
center should either be physically confined within the central regions
of the cluster or lie in the background. The radio galaxies located in
projection far from the cluster center, on the other
hand, obviously lie in the outer regions of the cluster. The same
conclusions can be drawn by considering the trend of the complexity of the
Faraday spectra of the radio galaxies with increasing projected
distance from the cluster center. The sources near the cluster center
are characterized by complex Faraday spectra, showing multiple peaks,
while the external radio galaxies have Faraday spectra with only one
peak.

The~~radio filaments~~of \object{A2255} show RM~~distributions ($<\rm RM> \sim
+20$~rad~m$^{-2}$, $\sigma_{\rm RM} \sim 10$~rad~m$^{-2}$) and Faraday
spectra similar to those of the external radio galaxies. This favors
the interpretation that these structures are not physically located in
the central regions of the cluster, but lie at the periphery. Their
elongated shape and high level of fractional polarization at high
frequency suggest that they are relics rather than part of the
halo. Their small RM-variance and Faraday depth are similar to those
of the external radio galaxies and of the Galactic foreground at the
location of the cluster. Therefore they are likely to be located in
the foreground of the cluster and not in the background.\\ The
filaments could be associated to merger shocks derived from the past
merger activity of the cluster. However, it is worth noting that the
information provided by their depolarization and RM distributions
suggests that the filaments should lie far away from the
cluster center, where shocks due to the LSS formation could also take
place
\citep{2003MNRAS.342.1009M,2003ApJ...585..128K}. That the
filaments are organized in a sort of ``net'' or ``web''
\citep{2008A&A...481L..91P} also
suggests that they could be relics associated with LSS-shocks.

\vspace{1.5cm}

\begin{acknowledgements}
 The Westerbork Synthesis Radio Telescope is operated by ASTRON
(Netherlands Institute for Radio Astronomy) with support from the
Netherlands Foundation for Scientific Research (NWO).
\end{acknowledgements}

\bibliographystyle{aa}
\bibliography{14158}
\end{document}